\documentclass[referee,pdflatex,sn-mathphys-num]{sn-jnl}

\usepackage{graphicx}%
\usepackage{multirow}%
\usepackage{amsmath,amssymb,amsfonts}%
\usepackage{amsthm}%
\usepackage{mathrsfs}%
\usepackage[title]{appendix}%
\usepackage{xcolor}%
\usepackage{textcomp}%
\usepackage{manyfoot}%
\usepackage{booktabs}%
\usepackage{algorithm}%
\usepackage{listings}%

\usepackage{bm}
\usepackage{bbm}
\usepackage[super]{nth}
\usepackage{placeins}

\newcommand{\mb}{\mathbf}

\usepackage{mathtools}

\DeclarePairedDelimiter{\abs} \lvert\rvert
\DeclarePairedDelimiter{\norm}\lVert\rVert

\usepackage{array}
\newcolumntype{L}{>{$}l<{$}} 
\newcolumntype{R}{>{$}r<{$}} 
\newcolumntype{C}{>{$}c<{$}} 
\usepackage{threeparttable}
\usepackage{makecell}
\usepackage{longtable}

\DeclareMathOperator*{\+}{\hphantom{-}}
\let\-\relax
\DeclareMathOperator*{\-}{{-}}

\usepackage{algpseudocodex}

\usepackage{subcaption}

\usepackage{siunitx}
\DeclareSIUnit\year{year}
\DeclareSIUnit{\au}{\astronomicalunit}
\DeclareSIUnit{\pc}{pc}
\DeclareSIUnit{\mas}{mas}
\DeclareSIUnit{\erg}{erg}
\DeclareSIUnit{\angstrom}{\text{\AA}}
\begin{document}

\title[Position and Time Determination without Prior State Knowledge via Onboard Optical Observations of Delta Scuti Variable Stars]{Position and Time Determination without Prior State Knowledge via Onboard Optical Observations of Delta Scuti Variable Stars}


\author*[1]{\fnm{Linyi} \sur{Hou}}\email{linyih2@illinois.edu}

\author[1]{\fnm{Ishaan} \sur{Bansal}\email{ibansal2@illinois.edu}}

\author[1]{\fnm{Clark} \sur{Davis}\email{clarkjd2@illinois.edu}}

\author[1]{\fnm{Siegfried} \sur{Eggl}}\email{eggl@illinois.edu}

\affil*[1]{\orgdiv{Department of Aerospace Engineering}, \orgname{University of Illinois Urbana-Champaign}, \orgaddress{\street{104 S Wright St.}, \city{Urbana}, \state{IL} \postcode{61820}}}


\abstract{We present a navigation concept for solving the lost in space and time problem using optical observations of $\delta$ Scuti variable stars. Only a small number of techniques exist that allow a spacecraft to recover from being lost in both space and time, which can be caused by a failure of the onboard clock and navigation systems. Optical observations of $\delta$ Scuti stars, which can be collected onboard from star trackers or navigation cameras, may enable autonomous position and time determination without requiring additional equipment or external communication. Our results indicate that less than one day of observation by the OSIRIS-APEX PolyCam may enable position and time determination accuracy within \qty{0.03}{\au} (3$\sigma$) and \qty{3}{\second} (3$\sigma$). }


\maketitle

\section{Introduction}\label{sec:introduction}

Onboard knowledge of spacecraft state can become lost or severely outdated due to power or data loss coupled with loss of contact to ground stations on Earth; this is known as the lost in space problem. In such cases, a spacecraft must autonomously take action to recover its state knowledge or reestablish communication. 

Most onboard spacecraft command and data handling (C\&DH) systems include contingencies for lost in space or similar scenarios. The New Horizons spacecraft has a ``Sun Acquisition'' mode in which its high-gain antenna is pointed to the Sun to send a signal indicating a critical fault if communication with the Earth cannot be established \cite{fountain2008new}. The Jupiter Icy Moons Explorer (JUICE) and Rosetta spacecraft both have an ``Earth Strobing'' mode in which the spacecraft rotates slowly with the medium gain antenna sweeping past Earth each rotation \cite{ecale2018juice,keil2007contingency}. However, such contingencies for the lost in space scenario ultimately rely on ground stations to provide commands that update the spacecraft state, which has proven difficult in the past. For example, the STEREO-Behind satellite was recovered briefly through extensive search operations conducted using the Deep Space Network (DSN) over 22 days \cite{cox2018fall}. More recently, Voyager 2 was inadvertently commanded to point its antenna away from the Earth and required signaling from the DSN to correct its attitude \cite{voyager2news}. 

Ground-based tracking and recovery methods rely on spacecraft either being able to receive signals regardless of antenna orientation or being able to point their antenna in the direction of Earth with sufficient accuracy. The former requires a powerful transmitter and proximity to Earth, while the latter may be challenging for interplanetary missions. Furthermore, for deep space missions, ground-based signaling typically relies on the DSN, whose schedule has become overloaded due to the growth in mission demand and data return \cite{johnston2020scheduling}. In contrast, techniques that allow a spacecraft to autonomously recover its position and time using onboard measurements alone are more scalable as the number of operational spacecraft increases. 

While numerous methods have been proposed to solve the lost in space problem using onboard measurements alone (assuming known time) \cite{tanygin2014closed,hollenberg2020geometric,hou2022xnav}, only a handful of methods are capable of recovering spacecraft that are lost in both space \textit{and time}. Adams and Peck introduced an optical navigation (OpNav)-based solution applicable within the vicinity of the Earth and Moon \cite{adams2017lost}. Dahir developed a method using star trackers to observe the Sun, Jupiter, and its moons for navigation between \qty{1}{\au} to \qty{5}{\au} from the Sun \cite{dahir2020lost}. Sala et al. devised a lost in space and time algorithm based on observations of X-ray pulsars \cite{sala2004feasibility}. 

Solutions to the lost in space and time problem that depend on signal sources from within the Solar System have several limitations. The location of those signal sources are highly sensitive to the current time and the observer's position, which means the observer may need to search the entire sky to find a signal. Additionally, illumination by the Sun and occlusion by nearby planets makes these techniques only applicable in certain parts of the Solar System (such as the vicinity of Earth or within \qty{5}{\au} of the Sun) \cite{broschart2019kinematic}. Signal sources from beyond the Solar System have smaller parallax from temporal and positional changes in the Solar System and can largely mitigate the aforementioned issues. However, the only existing lost in space and time navigation algorithms based on extrasolar sources utilize X-ray pulsars, which are faint and require long observation times with specialized detectors. Additionally, extensive computation is required to perform ambiguity resolution \cite{hou2022xnav}. 

In this paper, we present a novel solution for autonomous recovery from being lost in both space and time by using onboard observations of $\delta$ Scuti variable stars. The proposed method is similar in principle to X-ray pulsar navigation, making use of signal time of arrival (TOA) to determine the observer state. However, since $\delta$ Scuti variable stars can be observed in the optical band, no specialized X-ray detector is required to measure these signals. Our method may be performed almost anywhere in the Solar System using a single optical detector such as a star tracker or navigation camera. 

The drawbacks to the proposed navigation method include long observation times and poor accuracy. Simulation results suggest that observation arcs on the order of one day are required to achieve position and time accuracies on the order of \qty{1e-2}{\au} and \qty{10}{\second}, which is insufficient for a standalone navigation technique. We envision this method being used in conjunction with other higher accuracy navigation techniques that would otherwise be infeasible in a lost in space and time scenario. For example, the proposed method may be combined with on board planetary ephemerides to determine the relative position of the spacecraft with respect to the Earth to sufficient accuracy so as to enable re-acquisition of contact with ground stations.

\section{\texorpdfstring{$\delta$}{Delta} Scuti Variable Stars}\label{sec:dsct-variables}

\begin{figure}[!ht]
	\centering
	\includegraphics[width=0.95\linewidth]{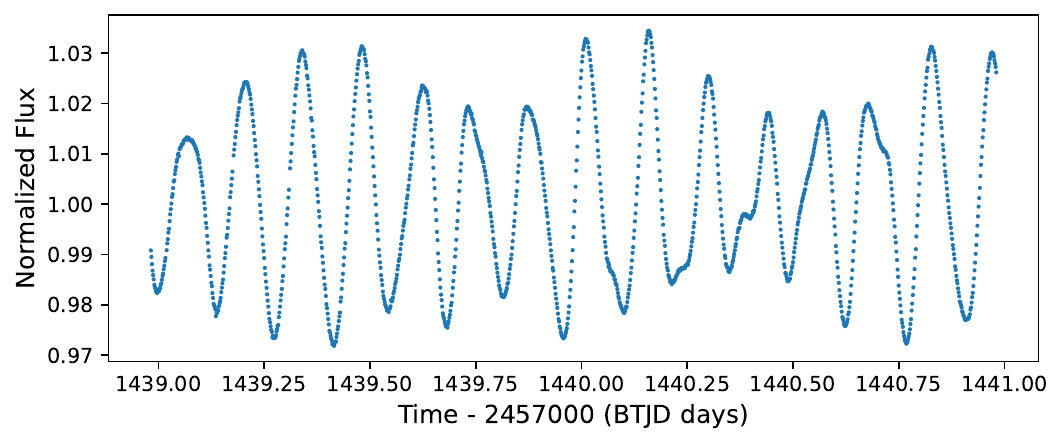}
	\caption{Normalized light curve of X Caeli.}
	\label{fig:example-light-curve}
\end{figure}

\begin{figure}[!ht]
	\centering
	\includegraphics[width=0.95\linewidth]{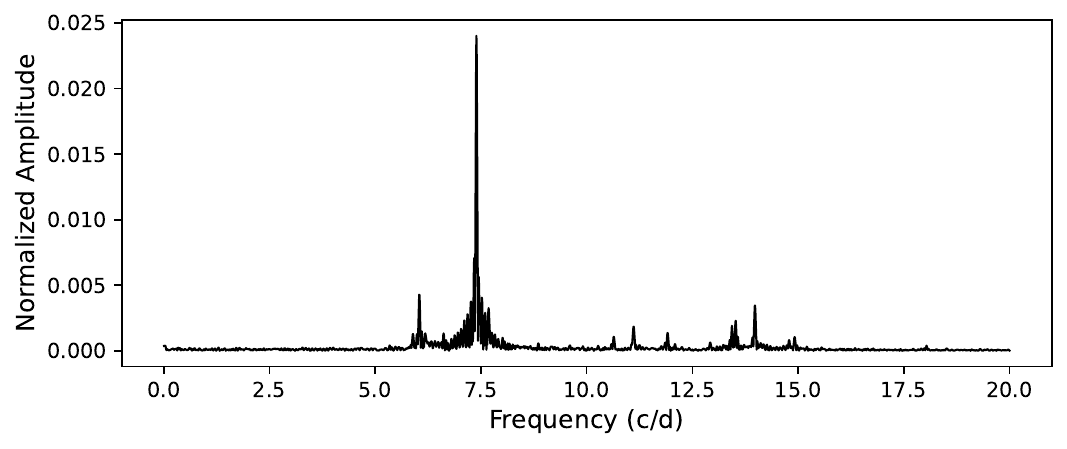}
	\caption{Pulsation spectra of X Caeli. The dominant frequencies --- 7.392, 6.046, and 13.980 cycles per day --- are in close agreement with results from Ref. \cite{mantegazza1999simultaneous}.}
	\label{fig:example-cycles-per-day}
\end{figure}

$\delta$ Scuti stars are a type of pulsating variable star whose luminosity varies due to the periodic expansion and contraction of the surface layers of the star \cite{eyer2008variable,aerts2010asteroseismology}. Each $\delta$ Scuti star's luminosity variation may comprise many pulsation modes with periods ranging from a few minutes to several hours \cite{handler2009delta,bowman2017amplitude,breger2009period}. Optical detectors such as star trackers and navigation cameras may be used to measure the varying flux of a $\delta$ Scuti star. Figures~\ref{fig:example-light-curve}-\ref{fig:example-cycles-per-day} show the light curve and dominant pulsation frequencies of the $\delta$ Scuti star X Caeli as measured by the Transiting Exoplanet Survey Satellite (TESS). 

\subsection{Light Curve Models}

The light curve of a $\delta$ Scuti star can be modeled as the sum of its pulsation modes, which in turn can each be modeled as sinusoids \cite{kovacs1989photoelectric}. More sophisticated modeling of individual pulsation modes via Fourier series \cite{poretti2001fourier} or seismic models \cite{zima2006new} exist but are not considered in this analysis. The model of a $\delta$ Scuti star's light curve is:
\begin{equation}
	\label{eqn:light-curve-ssb}
	m_\mb{b}(t) = A_0 + \sum_i A_i\cdot \sin\left(2\pi{}f_i(t-t_0)+\phi_i\right)
\end{equation}
where $A_i$, $f_i$, $\phi_i$ are the amplitude, frequency, and phase offset of each pulsation mode, $A_0$ is the magnitude offset, and $t_0$ is the model epoch. For a given signal TOA $t$, the light curve $m_\mb{b}(t)$ computes the predicted magnitude of the $\delta$ Scuti star at the reference observatory $\mb{b}$. The reference observatory is commonly chosen to be the Solar System barycenter (SSB).

\begin{figure}[!ht]
	\centering
	\includegraphics[width=0.65\linewidth,trim=200 150 200 160,clip]{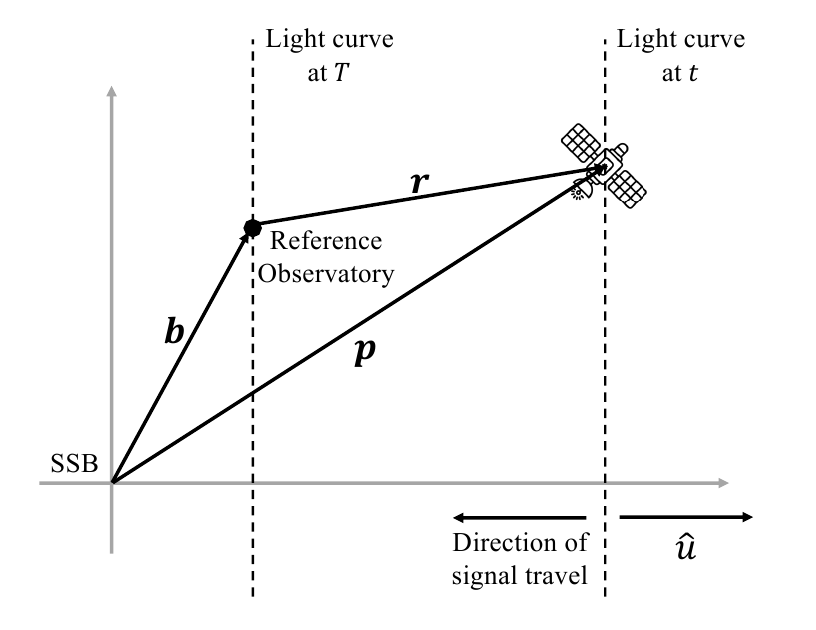}
	\caption{Diagram of measurement time transfer between a spacecraft and the reference observatory. Parallax and relativistic perturbations are neglected. A signal traveling along $-\hat{\mb{u}}$ can be measured at time $t$ at the spacecraft, and at time $T$ at the reference observatory. }
	\label{fig:geometry}
\end{figure}

The location of the observer affects the timing of the light curve model. For an observer at location $\mb{p}$, neglecting relativistic effects and stellar parallax, the same light curve becomes:
\begin{equation}
	\label{eqn:light-curve-observer}
	m_\mb{p}(t) = m_\mb{b}(t + \hat{\mb{u}}\cdot(\mb{p}-\mb{b})/c) = m_\mb{b}(T),
\end{equation}
where $c$ is the speed of light in a vacuum and $\hat{\mb{u}}$ is the unit line-of-sight (LOS) vector pointing from the observer to the variable star. This is a geometric time delay that can be illustrated by Figure~\ref{fig:geometry}, with $\mb{r} = \mb{p}-\mb{b}$. Consequently, measurements of the magnitude of $\delta$ Scuti variable stars contain information about the observer's position and time, which can be used to perform navigation.

It is assumed that the line of sight direction $\hat{\mb{u}}$ is fixed (no proper motion) and invariant for observers at different points in the Solar System. This validity of this assumption is justified in Appendix~\ref{appendix:los-assumption}. 

\subsection{Pulsation Stability Over Time}

The predictability of $\delta$ Scuti stars' light curves is a critical factor in selecting stars for navigation. Both the amplitude and period of $\delta$ Scuti stars' pulsation modes are known to change over time \cite{bowman2016amplitude,bowman2017amplitude,breger2009period,zhou2011period,boonyarak2011period}, which would invalidate light curve models stored onboard a spacecraft if not frequently updated. A study in 2016 found that 603 out of 983 $\delta$ Scuti stars exhibit significant amplitude variations in at least one pulsation mode \cite{bowman2016amplitude}. However, the remaining 380 $\delta$ Scuti stars did have stable pulsation mode amplitudes on the time scale of multiple years. Furthermore, variations in pulsation period are often well-modeled by a constant rate of change \cite{zhou2011period,boonyarak2011period}. A brief analysis on the stability of the $\delta$ Scuti stars is presented in Appendix~\ref{appendix:stability-analysis}.

While the present study does not explicitly identify a set of $\delta$ Scuti stars with stable amplitudes and periods suitable for navigation, it may be possible to establish a catalog of stable $\delta$ Scuti stars with these characteristics to avoid changes in the light curve models. For the remainder of the paper, it is assumed that the selected $\delta$ Scuti stars have stable periods and amplitudes over the observation period.

\FloatBarrier
\section{Navigation Algorithm}\label{sec:algorithm}

The lost in space and time navigation algorithm implemented in this paper determines the observer's position and time relative to a reference observatory. It is comprised of three parts: a gradient descent step to convert flux measurements/digital signals into signal TOAs, a linear system to solve for position and time given signal TOAs of many $\delta$ Scuti stars, and a search algorithm to resolve the ambiguity in the signal TOAs for each star. 

\subsection{Signal TOA Estimation}\label{sec:algorithm/signal-toa}

Suppose a spacecraft obtains a set of magnitude measurements $\{m_1,\cdots,m_n\}$ at $\{t_1',\cdots,t_n'\}$ according to the spacecraft clock. The spacecraft clock, measuring $t_k'$, is offset from the reference observatory clock (measuring $t_k$) by an unknown amount $t_\text{offset}$, such that: 
\begin{equation}
	\label{eqn:t_offset}
	t_k = t_k' + t_\text{offset}.
\end{equation}

We can find the reference observatory-centric TOAs of the measurements by minimizing $J$:
\begin{equation}
	\label{eqn:objective-fcn}
	J(\Delta{}t,K) = \dfrac{1}{n}\sum_1^n\Big(m_k-C\cdot{}m_\mb{b}(t_k'+\Delta{}t)\Big)^2
\end{equation}

$C$ is a scale factor that accounts for the difference between observed and predicted magnitudes caused by spectral responsivity differences between the detectors of the spacecraft and the observatory. $\Delta{}t$ represents the combination of $t_\text{offset}$ and the signal time of flight between the spacecraft and the reference observatory:
\begin{equation}
	\Delta{}t = t_\text{offset} + \hat{\mb{u}}\cdot\mb{r}/c
\end{equation}

As measured by the reference observatory clock, the time at which each measurement would arrive at the reference observatory is then:
\begin{equation}
	T_k = t_k' + \Delta{}t.
\end{equation}

To prevent aliasing in the light curve measurements, each $\delta$ Scuti star should be observed with a cadence of at least double its fastest pulsation mode. Measurements should also span at least one cycle of each star's highest-amplitude pulsation mode. These constraints generally mean that each $\delta$ Scuti star must be revisited each hour for several hours to obtain sufficient data for navigation. 

Due to the periodic nature of $\delta$ Scuti stars, there are an infinite number of possible solutions for $\Delta{}t$ corresponding to the local minima of $J$. This can be reduced to a finite amount by selecting only $\Delta{}t$ values for which $T_k$ is within a reasonable range based on the mission profile (e.g., all solutions within a few years of the spacecraft's launch date). The process of resolving the remaining ambiguity in the value of $\Delta{}t$ is very similar to the phase ambiguity problem encountered when performing X-ray pulsar navigation \cite{hou2022xnav} and is later addressed in Section~\ref{sec:algorithm/ambiguity-resolution}.

\subsection{Closed-form Solution without Ambiguity Resolution}

We first consider the closed-form navigation solution for when there is no ambiguity in the value of $\Delta{}t$ for any variable star. The formulation below assumes a fixed spacecraft position for all $\delta$ Scuti star observations, meaning its accuracy is limited by the amount of spacecraft motion during the observation period of at least several hours. In heliocentric orbit, this is on the order of \qty{0.01}{\au}. Analysis results from Appendix~\ref{appendix:stability-analysis} show that this is not a limiting factor, since signal TOA estimation accuracy | governed by the stability of $\delta$ Scuti stars' pulsation modes | will introduce errors of the same order of magnitude or greater (0.02-0.2 \qty{}{\au}).

Let there be $N$ $\delta$ Scuti stars with known positions on the celestial sphere whose LOS vectors are denoted by $\{\hat{\mb{u}}_1,\cdots,\hat{\mb{u}}_N\}$ in an inertial frame of reference such as the J2000 frame. Assume that a spacecraft whose coordinates are $\{\mb{p},t\}$ measures the light curve of star $i$, then generates estimate $\Delta{}t_i$. Following Eq.~\eqref{eqn:light-curve-observer} and Figure~\ref{fig:geometry}, the below equation must be satisfied: 
\begin{equation}
	\label{eqn:timing-eqn}
	\hat{\mb{u}}_i\cdot\mb{p} + c\cdot{}t = c \cdot{} T_i.
\end{equation}

Substituting $T = t' + \Delta{}t$ and $\mb{p} = \mb{r} + \mb{b}$, we have:
\begin{equation}
	\hat{\mb{u}}_i \cdot (\mb{r}+\mb{b}) + c \cdot t = c \cdot (t_i' + \Delta{}t_i)
\end{equation}

Separating unknowns and applying Eq.~\eqref{eqn:t_offset} gives:
\begin{equation}
	\hat{\mb{u}}_i \cdot \mb{r} + c \cdot t_\text{offset} = c \cdot \Delta{}t_i - \hat{\mb{u}}_i \cdot \mb{b}
\end{equation}

Combining observations from multiple $\delta$ Scuti stars into a linear system then yields Eq.~\eqref{eqn:Ax=d+n}, whose constituents are defined in Eq.~\eqref{eqn:matrix-definitions} and $\mb{n}$ is the noise vector in measuring $\mb{d}$ due to the uncertainty in $\Delta{}t_i$. Here, $\mb{A}\in\mathbb{R}^{N\times4}$; $\{\mb{s},\mb{d},\mb{n}\}\in\mathbb{R}^4$. 
\begin{equation}
	\label{eqn:Ax=d+n}
	\mb{A}\,\mb{s} = \mb{d} + \mb{n}
\end{equation}

\begin{equation}
	\label{eqn:matrix-definitions}
	\mb{A} = 
	\left[\mathbbm{1}_N,\left[
	\begin{array}{c}
		\hat{\mb{u}}_1^\top \\
		\vdots \\
		\hat{\mb{u}}_N^\top
	\end{array}\right]
	\right]
	\;,\;
	\mb{s} = 
	\left[\begin{array}{c}
		c\cdot{}t_\text{offset} \\
		\mb{r}
	\end{array}\right]
	\;,\;
	\mb{d} = 
	\left[\begin{array}{c}
		c\cdot{}\Delta{}t_1 - \hat{\mb{u}}_{_1}\cdot\mb{b} \\
		\vdots \\
		c\cdot{}\Delta{}t_N - \hat{\mb{u}}_{_N}\cdot\mb{b}
	\end{array}\right]
\end{equation}

Assuming that $\{\Delta{}t_1,\cdots,\Delta{}t_N\}$ have covariance matrix $\bm{\Omega}_{\Delta{}t}$, the covariance matrix of $\mb{d}$ is then $\bm{\Omega}_\mb{d} = c^2\bm{\Omega}_{\Delta{}t}$. Define $\mb{W} = \bm{\Omega}_\mb{d}^{-1}$. Here, $\{\mb{W}, \bm{\Omega}_{\Delta{}t},\bm{\Omega}_\mb{d}\}\in\mathbb{R}^{N\times{}N}$. The closed-form weighted least-squares solution to Eq.~\eqref{eqn:Ax=d+n} is then:
\begin{equation}
	\label{eqn:weighted_lsq}
	\mb{s}^* = \left(\mb{A}^\top\mb{W}\mb{A}\right)^{-1}\mb{A}^\top\mb{W} \mb{d}.
\end{equation}

The spacecraft's position relative to the SSB is then retrieved by $\mb{p} = \mb{r} + \mb{b}$, and the spacecraft clock can be synchronized with the reference observatory clock via $t = t' + t_\text{offset}$. Eq.~\eqref{eqn:weighted_lsq} represents a single solution for the observer's position and time given a set of solutions of $\Delta{}t_i$ for each $\delta$ Scuti star. Multiple possible values of $\Delta{}t_i$ would therefore produce many such solutions --- a method for resolving the ambiguity in $\Delta{}t_i$ is presented below.

\subsection{Search Algorithm for Ambiguity Resolution}\label{sec:algorithm/ambiguity-resolution}

\begin{figure}[ht!]
	\centering
	\includegraphics[width=0.8\linewidth,clip,trim=200 160 30 30]{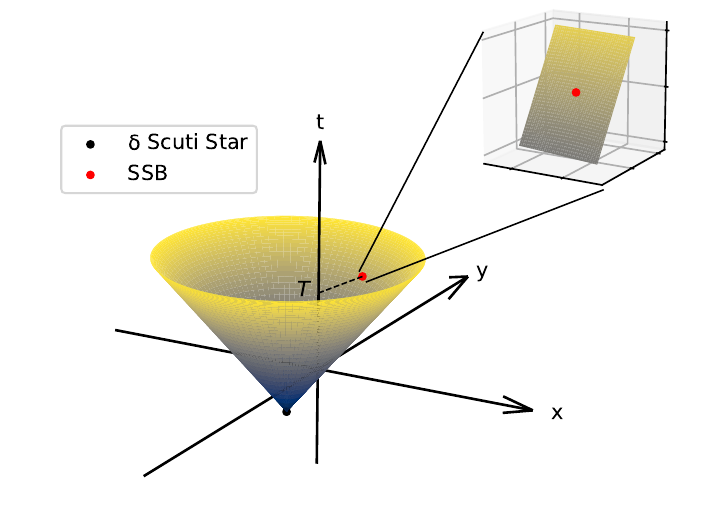}
	\caption{Illustration of a three-dimensional light cone originating from a $\delta$ Scuti star, representing the set of possible spacecraft coordinates given a signal measurement known to pass through the SSB at time $T$.}
	\label{fig:spacetime-signal}
\end{figure}

We utilize the uniqueness of overlapping signals from many $\delta$ Scuti stars to resolve ambiguity. Suppose that a spacecraft receives a signal from a $\delta$ Scuti star, and at that moment the unknown spacetime coordinate of the spacecraft is $\{\mb{p},t\}$. If the signal would have arrived at the SSB at time $T$ based on the estimation process outlined on the previous section, then $\{\mb{x},t\}$ and $\{\mb{b},T\}$ must be on the same light cone with slope $c$ whose vertex is at the $\delta$ Scuti star. $c$ is again the speed of light. Consequently, the set of possible spacetime coordinates for the spacecraft is reduced to the surface of light cone; this concept is illustrated for two-dimensional space in Figure~\ref{fig:spacetime-signal}.

For navigation within the Solar System where the length scales involved are far smaller than the distances between the Sun and any $\delta$ Scuti star, only a small portion of the light cone around the SSB is relevant. The curvature of the light cone is negligible at the scale of the Solar System, so the set of possible spacecraft spacetime coordinates forms a hyperplane. This is depicted in the enlarged view of the light cone in Figure~\ref{fig:spacetime-signal}. In similar literature concerning pulsar navigation, these hyperplanes are referred to as ``wavefronts'', so we will follow the same convention henceforth \cite{sala2004feasibility,sheikh2011spacecraft,lohan2021methodology}. 

\begin{figure}[!ht]
	\centering
	\begin{subfigure}[t]{0.32\linewidth}
		\centering
		\includegraphics[width=\linewidth,clip,trim={140 180 220 100}]{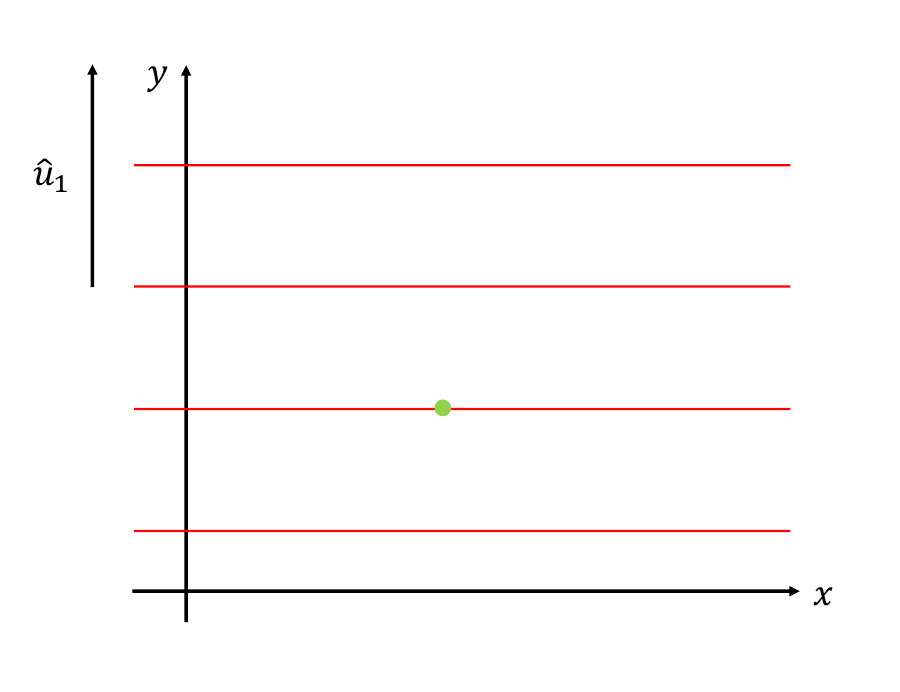}
		\caption{}
	\end{subfigure}
	\hfill
	\begin{subfigure}[t]{0.32\linewidth}
		\centering
		\includegraphics[width=\linewidth,clip,trim={140 180 220 100}]{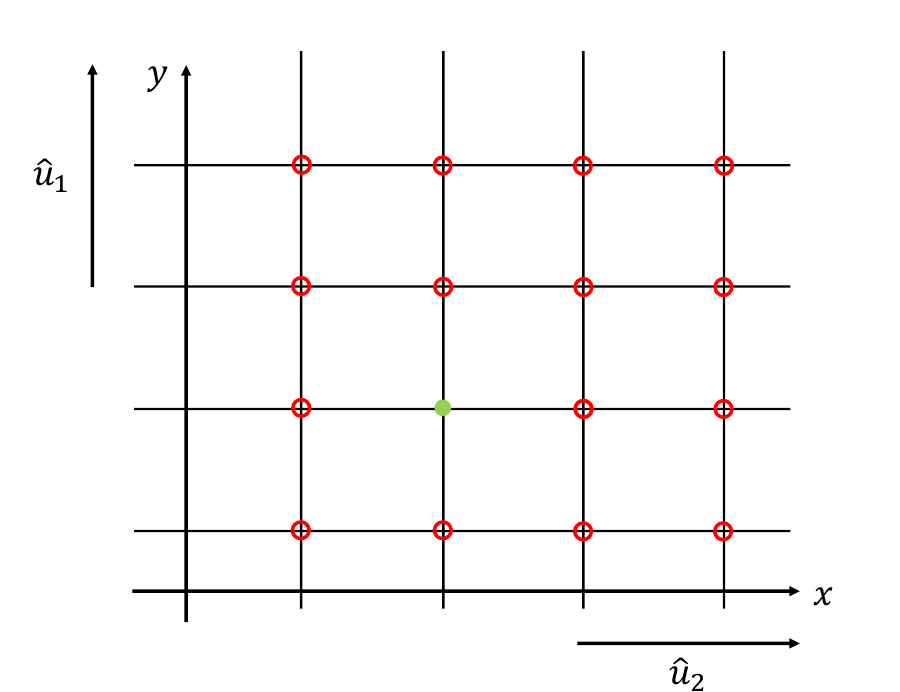}
		\caption{}
	\end{subfigure}
	\hfill
	\begin{subfigure}[t]{0.32\linewidth}
		\centering
		\includegraphics[width=\linewidth,clip,trim={140 180 220 100}]{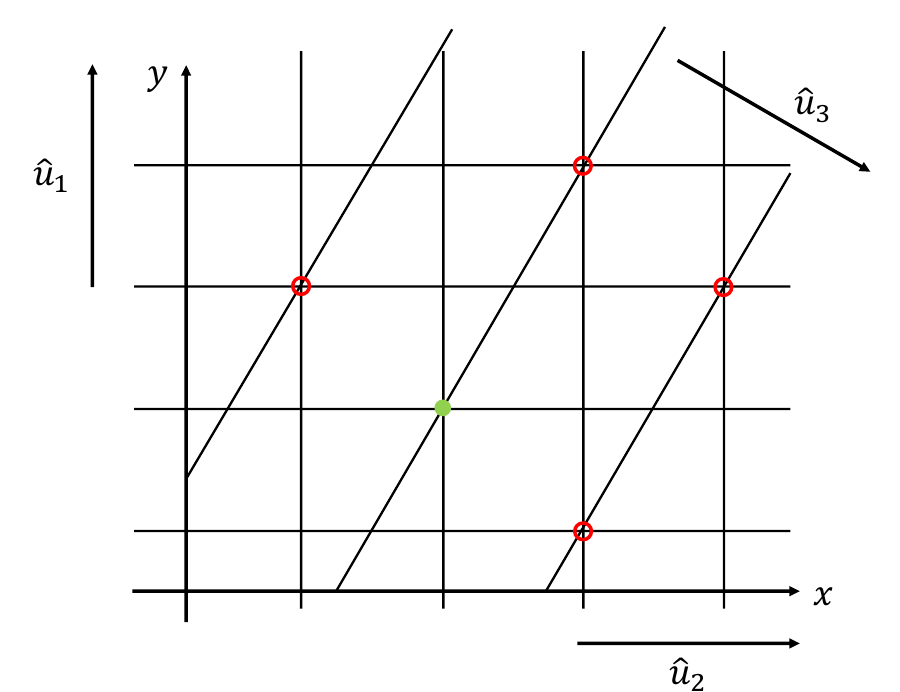}
		\caption{}
	\end{subfigure}
	\caption{Two-dimensional example of wavefront intersection. The $\delta$ Scuti star lines of sight are indicated by $\hat{\mb{u}}$. The observer's true spacetime coordinate is in green. The observer's spacetime coordinate can be on an infinite number of wavefronts --- shown in red --- when only observing the light curve of one $\delta$ Scuti star (a). Adding wavefronts from observations of other $\delta$ Scuti stars reduces set of feasible spacetime coordinates to intersections of wavefronts as shown in (b)-(c) \cite{lohan2021methodology}.}
	\label{fig:wavefront-definition}
\end{figure}

Each wavefront represents the possible spacecraft coordinates corresponding to one estimate of $\Delta{}t$ for a $\delta$ Scuti star. As previously discussed, there are an infinite number of local minima leading to an infinite number of estimates for $\Delta{}t$ per star. This ambiguity is resolved by combining observations of multiple $\delta$ Scuti stars, since the coordinate of the spacecraft must be at the intersection of the sets of wavefronts. Within a finite spacetime search region, the number of intersections can be reduced to one by observing a large enough number of $\delta$ Scuti stars. This concept is illustrated in two-dimensional space by Figure~\ref{fig:wavefront-definition}, where each wavefront is a line.

Hou presented an algorithm for ambiguity resolution in the context of X-ray pulsar navigation, which allows the concept in Figure~\ref{fig:wavefront-definition} to be performed efficiently \cite{hou2022xnav}. Given a convex, three-dimensional spatial search region, the algorithm updates a position solution by incrementally incorporating wavefronts from more pulsars. In this paper, the algorithm is adapted to $\delta$ Scuti stars in four-dimensional spacetime and summarized in Algorithm~\ref{alg:recursive_search}. $\sigma_{\Delta{}t}$ is the uncertainty in the estimate of $\Delta{}t$ from solving Eq.~\eqref{eqn:objective-fcn}. 

\begin{algorithm}[!ht]
	\caption{An ambiguity resolution algorithm for identifying spacecraft time and position from $\delta$ Scuti star light curve observations.}
	\label{alg:recursive_search}
	\begin{algorithmic}[1]
		\Function{candidate\_search}{$i$, $\tilde{\mb{s}}$, $\mb{d}$}
		\State $\{c\cdot\tilde{t}_\text{offset},\; \tilde{\mb{r}}\} \gets \tilde{\mb{s}}$
		\State $\bm{\tau}_i \gets \{ \Delta{}t \;|\; \nabla{}J(\Delta{}t,C) = 0,\; \nabla^2J(\Delta{}t)\succ 0\}$
		\Comment{Find possible $\Delta{}t$ values.}
		\If{$i \geq 4$}
			\Comment{Keep wavefronts near the existing solution.}
			\State $\bm{\tau}_i \gets \{ \Delta{}t \in \bm{\tau}_i\;|\; \abs{\hat{\mb{u}}_i\cdot\tilde{\mb{r}} + c\cdot{}\tilde{t}_\text{offset} - c\cdot\Delta{}t+\hat{\mb{u}}_i\cdot{\mb{b}}} < 3\sigma_{\Delta{}t}\}$
		\EndIf
		\Loop
			\State $\Delta{}t_i \gets$ \Call{next}{$\bm{\tau}_i$}
			\Comment{Iterate through each value in $\bm{\tau}_i$.}
			\State $d_i \gets c\cdot{}\Delta{}t_i - \hat{\mb{u}}_i\cdot\mb{b}$
			\If{$i \geq 4$}
				\LComment{$A_{ii}$ is full rank, compute weighted least-squares solution.}
				\State $\mb{A}_{ii} \gets \mb{A}[1:i,1:i]$
				\State $\mb{W}_{ii} \gets \mb{W}[1:i,1:i]$
				\State $\tilde{\mb{s}} \gets \left(\mb{A}_{ii}^\top\mb{W}_{ii}\mb{A}_{ii}\right)^{-1}\mb{A}_{ii}^\top\mb{W}_{ii}\mb{d}[1:i]$
				\If{$\tilde{\mb{s}}$ outside search domain} \textbf{continue}
				\EndIf
				\If{$\mb{A}_{ii}\tilde{\mb{s}}-\mb{d}[1:i]$ exceeds error bounds} \textbf{continue}
				\EndIf
			\EndIf
			\If{$i < N$}
			\LComment{Include more $\delta$ Scuti stars.}
			\State \Call{candidate\_search}{$i+1$, $\tilde{\mb{s}}$, $\mb{d}$}
			\Else
			\State store $\tilde{\mb{s}}$ \Comment{All $\delta$ Scuti star observations used. Record valid solution.}
			\EndIf
		\EndLoop
		\EndFunction
	\end{algorithmic}
\end{algorithm}

The algorithm assumes a convex four-dimensional search region. If the spacecraft state and uncertainty is known for a prior time, it is possible to propagate the spacecraft state covariance to generate the search region. If the spacecraft state is unknown, the search region can simply be the entire Solar System. A non-convex region can be partitioned into several smaller, convex regions for which the algorithm can be applied. Nonlinear dynamics such as proper motion, the parallax effect, and wavefront distortions due to relativistic effects are omitted because their impact on timing accuracy is negligible compared to measurement uncertainty. In other words, it is assumed that the wavefronts are planar and parallel.

\FloatBarrier
\section{Simulated Observations}\label{sec:simulated-observations}

A navigation case study is constructed by simulating the flux measurement of $\delta$ Scuti stars as measured by optical instruments onboard OSIRIS-APEX during its cruise to the asteroid Apophis. In this section, the methodology for simulating $\delta$ Scuti star observations is described.

\subsection{Spacecraft Trajectory}

\begin{figure}[ht!]
	\centering
	\includegraphics[width=0.85\linewidth,trim=150 65 120 95,clip]{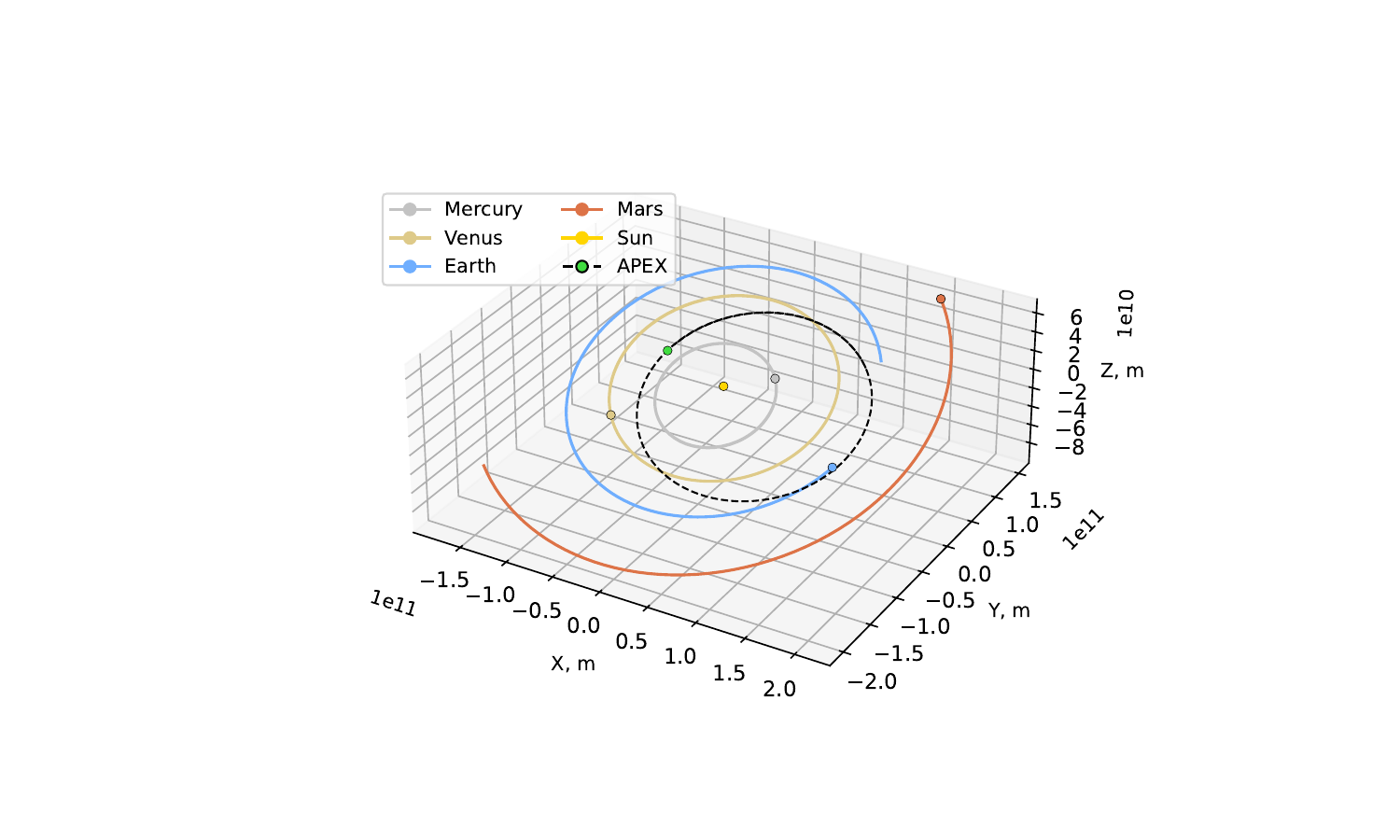}
	\caption{The positions of the inner planets and OSIRIS-APEX are shown for MJD 60555 (Sep. 2, 2024). OSIRIS-APEX and the Earth are in opposition.}
	\label{fig:ephemerides}
\end{figure}

OSIRIS-APEX will undergo several perihelion passes on its cruise to encounter asteroid Apophis in 2029, during which time communication with the Earth may not be available \cite{dellagiustina2023osiris}. The ability to autonomously recover from being lost in space and time may be especially important in these scenarios. This work simulates $\delta$ Scuti star observations made by OSIRIS-APEX on September 2, 2024 during a perihelion passage. The trajectory and position of the four inner planets are shown in Figure~\ref{fig:ephemerides}.
\FloatBarrier

\subsection{Detector Measurements}

PolyCam and MapCam from the OSIRIS-REx Camera Suite (OCAMS) are considered for the present analysis. Sensor parameters, which are shared for all cameras in OCAMS, are shown in Table~\ref{tab:orex-sensors}. The focal length ($f$) and aperture size ($D$) of PolyCam are \qty{629}{\mm} and \qty{175}{\mm} respectively. For MapCam, they are \qty{125}{\mm} and \qty{38}{\mm}.

\begin{table}[!ht]
	\centering
	\caption{OCAMS sensor specifications \cite{rizk2018ocams,golish2020ground}.}
	\label{tab:orex-sensors}
	\begin{tabular}{ccccc}
		\toprule
		\makecell{Pixel Size\\$A$ (\si{\micro\meter^2})} &
		\makecell{Readout Noise\\$RN$ (e$^{\-}$)} &
		\makecell{Dark Current\\$d$ (\si{\pico\ampere\per\cm^2})} &
		\makecell{Quantum\\Efficiency} &
		\makecell{Gain, $G$\\(e$^{\-}$/ADU)} \\
		\midrule
		$6.5\times8.5$ & $<50$ & 0.065 & 0.3-0.4 & 4.5 \\
		\bottomrule
	\end{tabular}
\end{table}

We simulate the digital signal measured by the detector. First, the magnitude $m$ of a $\delta$ Scuti star magnitude is converted to photon flux density $\Phi$ using Eq.~\eqref{eqn:mag-to-photon-flux}. $f_\lambda = \qty{3.631e-9}{\erg\per\cm^2\per\second\per\angstrom}$ is the Johnson/Bessell V-band zero point flux in the Vega magnitude system and $\Delta\lambda_\text{eff} = \qty{88}{\nm}$ is the effective bandwidth. $Q_{ph} = \frac{h\cdot{}c}{\lambda_\text{avg}}$ is the energy of one photon at the average wavelength $\lambda_\text{avg}$ of the V-band, where $h$ is Planck's constant and $c$ is the speed of light. The filter is assumed to be 100\% efficient.
\begin{equation}
	\label{eqn:mag-to-photon-flux}
	\Phi = \dfrac{f_\lambda \times \Delta\lambda_\text{eff} \times 10^{-m/2.5}}{Q_{ph}}
\end{equation}

Given an exposure time of $t$ and aperture area $A_\text{det} = \pi{}(D/2)^2$, the number of source photons collected by the detector is:
\begin{equation}
	\label{eqn:photon-count}
	S_{ph} = \text{round}\big(\Phi \times A_\text{det} \times t\big)
\end{equation}

The captured source photons are converted to electrons based on the quantum efficiency (QE) of the sensor via Eq.~\eqref{eqn:photons-to-electrons}. While QE is typically a function of wavelength, we assume a scalar value to represent the average QE across the V-band for simplicity. 
\begin{equation}
	\label{eqn:photons-to-electrons}
	S_{e^-} = S_{ph} \times QE
\end{equation}

The source signal is obtained by converting the analog signal to a digital signal:
\begin{equation}
	S = S_{e^-} / G
\end{equation}

The sky background flux is assumed to be \qty{21.5}{mag\per arcsec^2}. The line-spread of the OCAMS suite at full-width-half-maximum is under 2 pixels, so we assume that the aperture mask comprises $\text{npix} = 3\times3$ pixels. The digital sky background signal is:
\begin{equation}
	\label{eqn:sky-background}
	S_s = \text{round}\bigg(\dfrac{f_\lambda \times \Delta\lambda_\text{eff} \times 10^{-m_\text{sky}/2.5}}{Q_{ph}} \times A \times f^{-2} \times \text{npix} \times t \bigg) \times QE / G
\end{equation}

The signal from dark current is calculated as follows:
\begin{equation}
	S_d = \text{round}(d \times A \times \text{npix} \times t) \times QE / G
\end{equation}

Finally, shot noise from the above sources are simulated by sampling from a Poisson distribution $\mathcal{P}(\cdot)$ to obtain the final digital signal:
\begin{equation}
	S_\text{tot} = \mathcal{P}(S + S_s + S_d).
\end{equation}

Changes in the spacecraft position between measurements, as well as light propagation delay between the SSB and the spacecraft, are accounted for. However, stellar aberration, lens distortion, and other effects that impact the apparent position of the star on the imaging sensor are not considered.

\subsection{Star Selection and Light Curve Generation}

\begin{figure}[ht!]
	\centering
	\includegraphics[width=\linewidth]{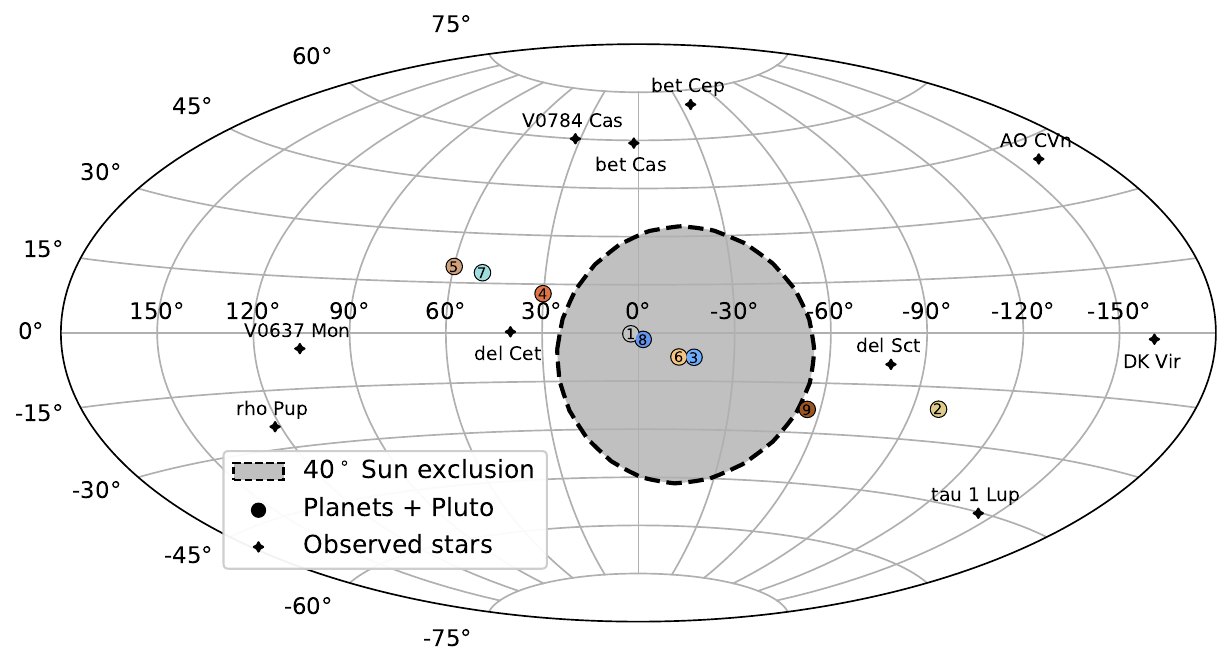}
	\caption{Distribution of 10 $\delta$ Scuti stars and Solar System planets from the view of OSIRIS-APEX used in the analysis are shown in the J2000 equatorial coordinate system using Aitoff projection.}
	\label{fig:aitoff}
\end{figure}

From the International Variable Star Index (VSX) \cite{watson2006international}, we compiled a list of $\delta$ Scuti stars with apparent magnitudes of $m_V < 7.0$ in the Johnson $V$ filter, brightness modulation amplitudes $\Delta{V} > 0.04 \text{ mag}$, and dominant frequencies $f > 5.0 \text{ cycles/day}$. The selection criteria was chosen to ensure detectable levels of pulsation within the capabilities of MapCam and PolyCam. A total of 66 $\delta$ Scuti stars met the above criteria and are listed in the appendix, Table~\ref{tab:dsct-table}. While the present analysis exclusively considers $\delta$ Scuti stars, the concept may extend to other optical variable stars such as RR Lyrae and Cepheid variables. A subset of 10 $\delta$ Scuti stars were arbitrarily selected to be observed. They are shown in Figure~\ref{fig:aitoff} along with the Solar System planets. 

\begin{figure}[ht!]
	\centering
	\includegraphics[width=\linewidth]{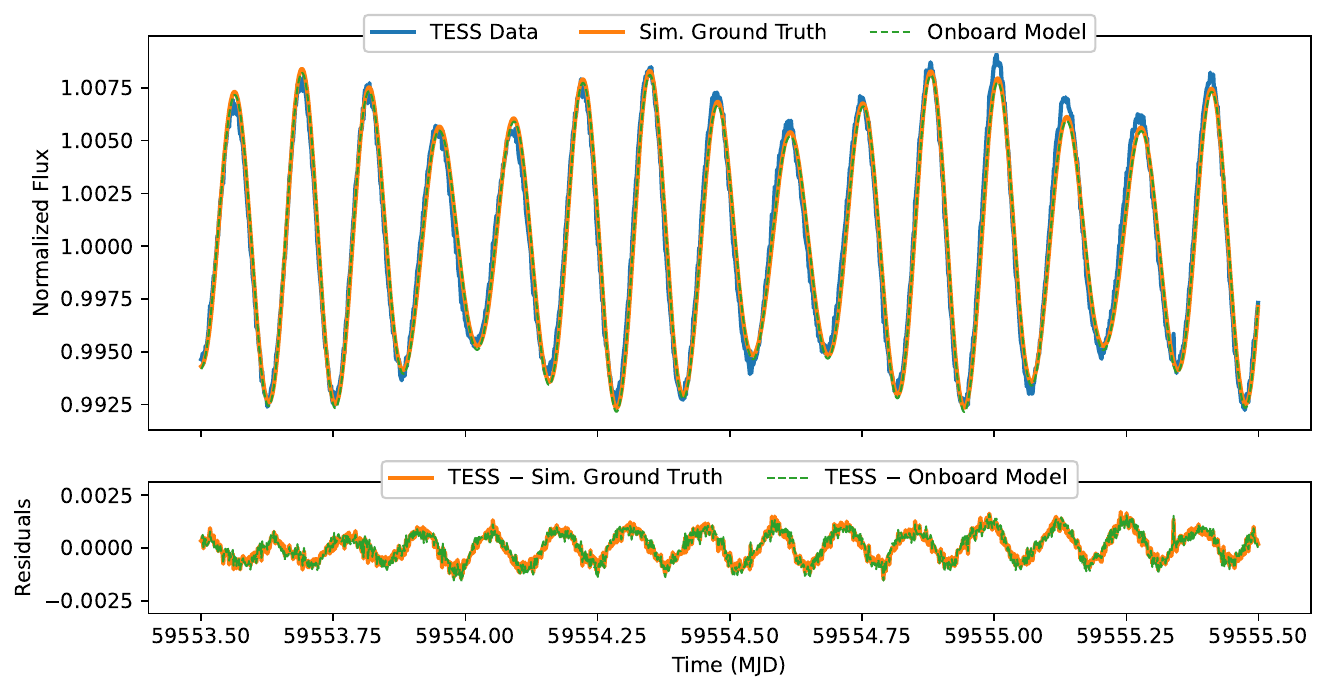}
	\caption{TESS observation data of DK Virginis is used to reconstruct its pulsation model with 1\% and 5\% amplitude cutoffs (top panel). The model fit residuals are shown (bottom panel).}
	\label{fig:example-lc-model}
\end{figure}

The light curves of $\delta$ Scuti stars are generated using observation data from TESS. First, Lomb-Scargle periodograms are used to identify the frequency and amplitude of pulsation modes with signal-to-noise ratios (SNR) greater than 3. Then, a curve fit is performed to determine the phase of each mode. Pulsation modes with amplitudes smaller than 1\% that of the highest amplitude mode are discarded, and the resulting light curve model is considered the simulation ground truth. To simulate onboard data containing light curve model inaccuracies, a lower-fidelity light curve is generated by discarding modes with amplitudes less than 5\% that of the highest amplitude mode prior to performing the curve fit. Figure~\ref{fig:example-lc-model} shows an example of the resulting light curve models for DK Virginis.

\begin{figure}[ht!]
	\centering
	\begin{subfigure}{0.49\linewidth}
		\centering
		\includegraphics[width=\linewidth]{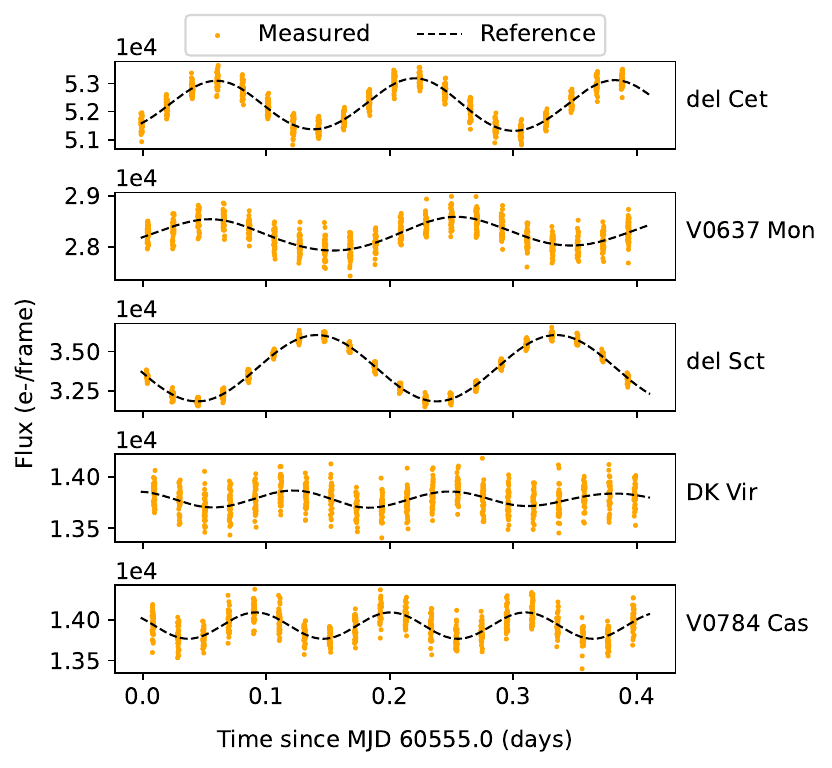}
	\end{subfigure}
	\hfill
	\begin{subfigure}{0.49\linewidth}
		\centering
		\includegraphics[width=\linewidth]{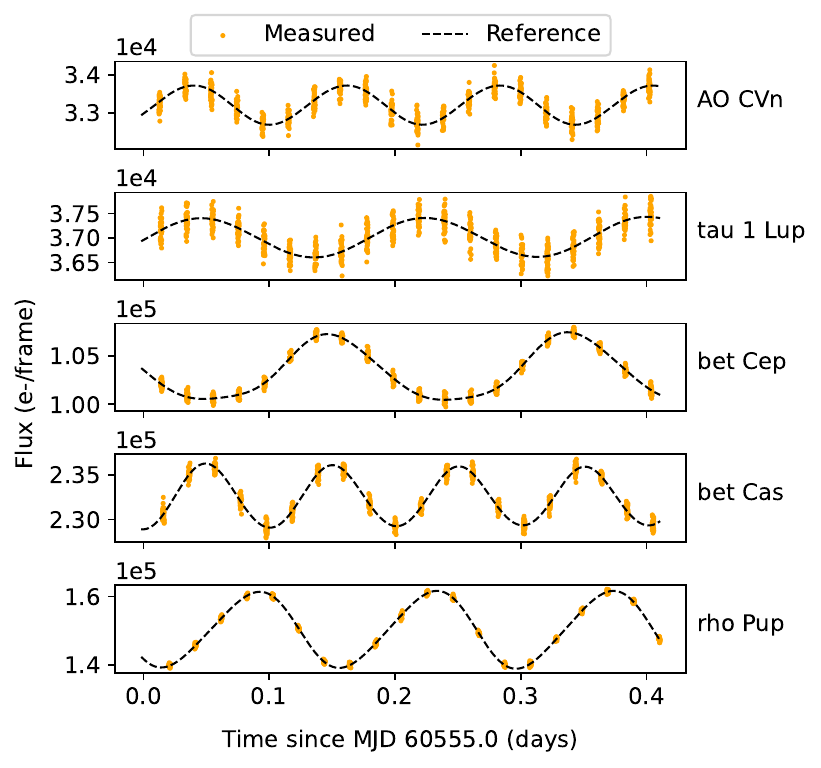}
	\end{subfigure}
	\caption{Ground truth light curves and simulated flux measurements by MapCam.}
	\label{fig:simulated-observations}
\end{figure}

It is assumed that the spacecraft will sequentially observe each $\delta$ Scuti star for 2 minutes and then revisit each star in the same order, repeating the process 20 times. Images are taken at a \qty{3}{\second} cadence; a \qty{1}{\minute} slew time is allotted between observations of different stars. The ground truth light curves and simulated flux measurements by MapCam are shown in Figure~\ref{fig:simulated-observations}.

\section{Feasibility Assessment}\label{sec:feasibility}

We performed error analysis for the proposed navigation algorithm using Monte Carlo simulation methods. First, the nominal case is presented for both MapCam and PolyCam. Then, the effects of varying observation conditions are explored. For all simulations, we assume that the position of the spacecraft is known to be contained within a three-dimensional sphere of radius \qty{40}{\au}; the current time is known to be contained within a range of $\pm{}10$ days; the attitude of the spacecraft is assumed to be known through star tracker observations.

\subsection{Nominal Case}\label{sec:feasibility/nominal}

\begin{figure}[!ht]
	\centering
	\includegraphics[width=\linewidth]{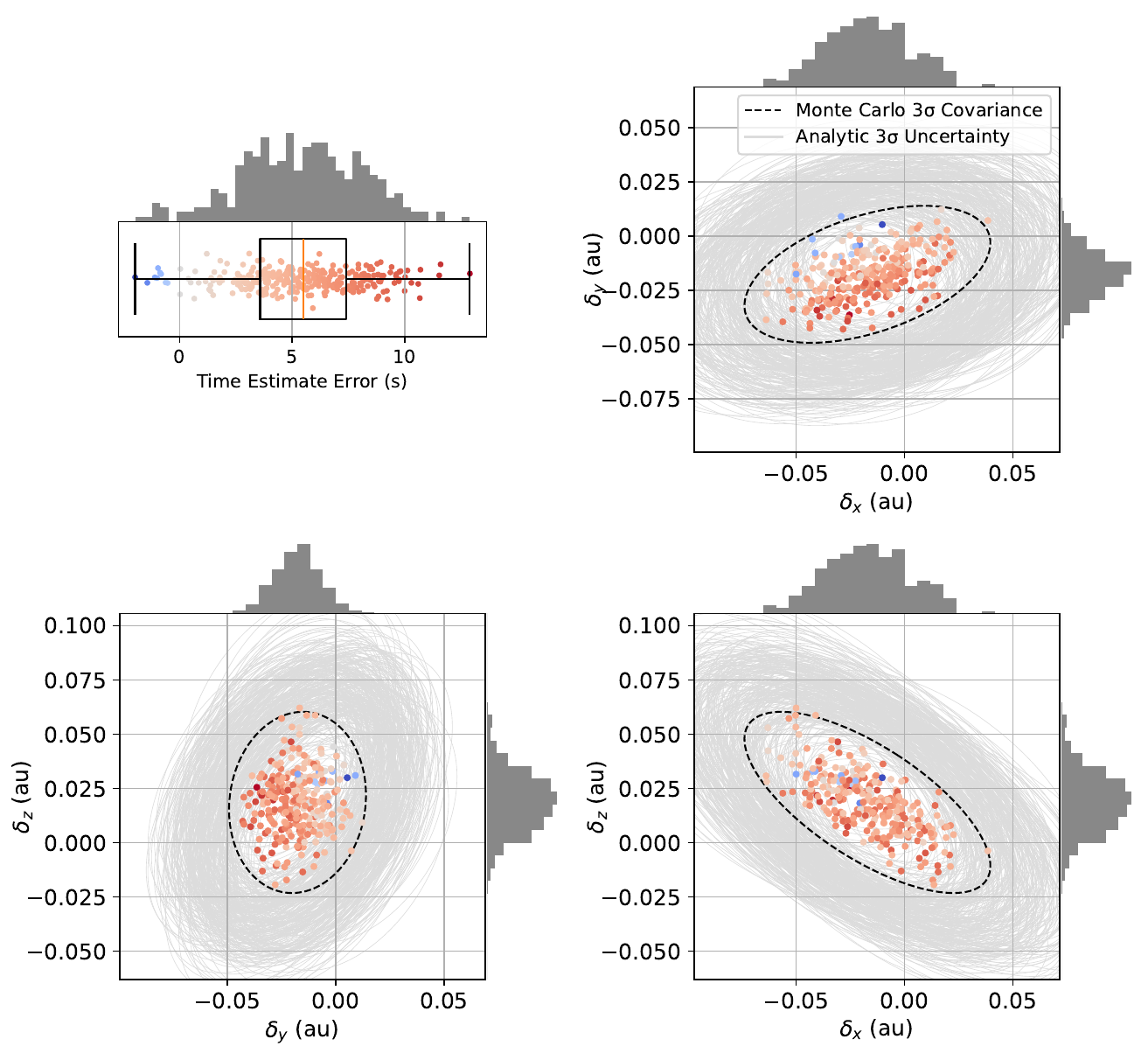}
	\caption{Time and position error distribution for a 300-sample Monte Carlo simulation of the proposed navigation method using MapCam in the nominal case. The numerical and analytic $3\sigma$ uncertainty ellipses are shown. }
	\label{fig:nominal-mapcam}
\end{figure}

Figure~\ref{fig:nominal-mapcam} shows the result of a 300-sample Monte Carlo simulation using MapCam with observation conditions described in Section~\ref{sec:simulated-observations}. The time estimate error has mean \qty{5.40}{\second} and standard deviation \qty{2.67}{\second}. The position estimate error has mean \qty{3.08e-2}{\au} and standard deviation \qty{2.57e-2}{\au}. All 300 Monte Carlo instances succeeded in finding the spacecraft coordinate.

\begin{figure}[!ht]
	\centering
	\includegraphics[width=\linewidth]{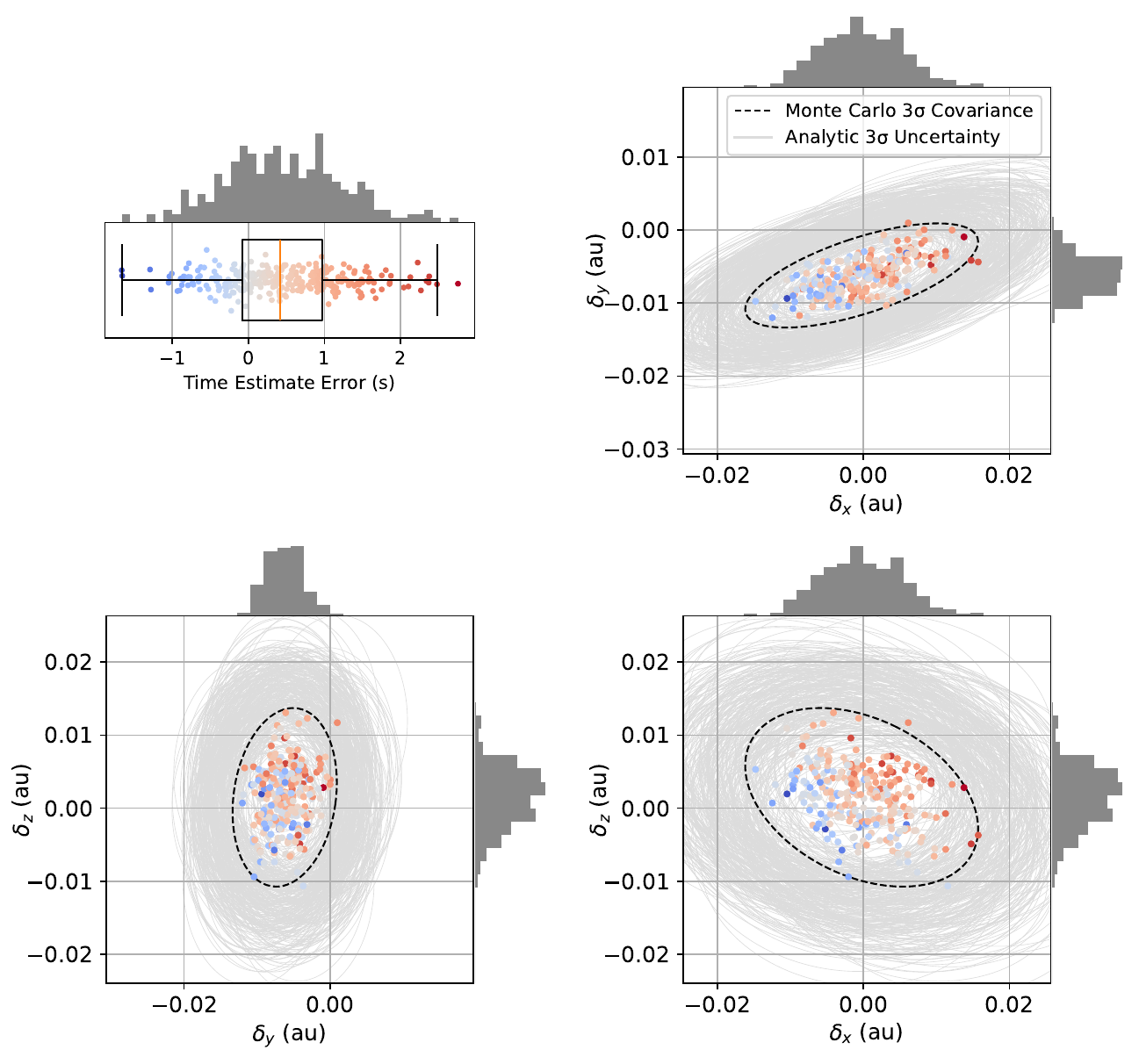}
	\caption{Time and position error distribution for a 300-sample Monte Carlo simulation of the proposed navigation method using PolyCam in the nominal case. The numerical and analytic $3\sigma$ uncertainty ellipses are shown. }
	\label{fig:nominal-polycam}
\end{figure}

The same analysis is performed for simulated observations made with PolyCam, with results shown in Figure~\ref{fig:nominal-polycam}. The time estimate error has mean \qty{0.46}{\second} ($1\sigma=\qty{0.78}{\second}$). The position estimate error has mean \qty{0.64e-2}{\second} ($1\sigma=\qty{0.71e-2}{\second}$). All 300 Monte Carlo instances succeeded in finding the spacecraft coordinate.

\FloatBarrier
It should be noted that our method can accommodate search regions of far greater or smaller sizes, but more $\delta$ Scuti stars may need to be observed in order to resolve ambiguity in a larger search space. Additionally, the compute time increases linearly with respect to the four-dimensional volume of the search region. For reference, solving 300 Monte Carlo instances takes less than 30 seconds on an Apple M2 CPU (single-threaded performance).

\subsection{Parameterized Study}

The effect of changing observation parameters on the position and time error of the proposed navigation method is analyzed using Monte Carlo simulation, with results shown in Tables~\ref{tab:parameter-study-mapcam}-\ref{tab:parameter-study-polycam}. Parameters considered include the observation time spent on each star, the exposure time for each observation, and the number of revisits | meaning how many times the sequential observation of all stars is repeated. In both tables, the corresponding nominal scenario discussed in Section~\ref{sec:feasibility/nominal} is shown as Case 1. 

\newcommand{\bu}[1]{\underline{\textbf{#1}}}
\begin{table}[ht!]
	\centering
	\caption{Position and time solution error and uncertainty for varied observation parameters using MapCam.}
	\label{tab:parameter-study-mapcam}
	\begin{tabular}{ccccccc}
		\toprule
		Case &
		\makecell{Observation\\Time (s)} &
		\makecell{Exposure\\Time (s)} &
		Revisits &
		\makecell{Position\\Error($1\sigma$) (au)} &
		\makecell{Time\\Error($1\sigma$) (s)} &
		\makecell{\# of Cases\\Solved} \\
		\midrule
		1 & 120 & 3.0 & 20 & $3.08(2.57)\times10^{-2}$ & $\+5.40(2.67)$ & 300 \\
		2 & 240 & 3.0 & 10 & $4.07(2.59)\times10^{-2}$ & $\-1.47(2.95)$ & 300 \\
		3 & 480 & 3.0 & 5 & $28.4(4.18)\times10^{-2}$ & $\-77.4(9.40)$ & 257 \\
		4 & 120 & 2.0 & 20 & $2.64(3.42)\times10^{-2}$ & $\+7.87(3.64)$ & 300 \\
		5 & 120 & 1.0 & 20 & $7.54(5.91)\times10^{-2}$ & $\+17.6(8.44)$ & 300 \\
		6 & 120 & 3.0 & 10 & $9.34(3.98)\times10^{-2}$ & $\-12.9(4.42)$ & 300 \\
		7 & 120 & 3.0 & 30 & $3.34(2.05)\times10^{-2}$ & $\+1.83(2.19)$ & 300 \\
		\bottomrule
	\end{tabular}
\end{table}

\begin{table}[ht!]
	\centering
	\caption{Position and time solution error and uncertainty for varied observation parameters using PolyCam.}
	\label{tab:parameter-study-polycam}
	\begin{tabular}{ccccccc}
		\toprule
		Case &
		\makecell{Observation\\Time (s)} &
		\makecell{Exposure\\Time (s)} &
		Revisits &
		\makecell{Position\\Error($1\sigma$) (au)} &
		\makecell{Time\\Error($1\sigma$) (s)} &
		\makecell{\# of Cases\\Solved} \\
		\midrule
		1 & 120 & 3.0 & 20 & $0.64(0.71)\times10^{-2}$ & $\+0.46(0.78)$ & 300 \\
		2 & 240 & 3.0 & 10 & $2.36(0.68)\times10^{-2}$ & $\-2.97(0.77)$ & 300 \\
		3 & 480 & 3.0 & 5 & $5.77(0.87)\times10^{-2}$ & $\-16.3(1.31)$ & 269 \\
		4 & 120 & 2.0 & 20 & $0.79(0.85)\times10^{-2}$ & $\-0.02(0.89)$ & 300 \\
		5 & 120 & 1.0 & 20 & $0.78(1.04)\times10^{-2}$ & $\+1.05(1.04)$ & 300 \\
		6 & 120 & 3.0 & 10 & $3.86(1.01)\times10^{-2}$ & $\-12.8(1.13)$ & 300 \\
		7 & 120 & 3.0 & 30 & $1.09(0.56)\times10^{-2}$ & $\+0.08(0.65)$ & 300 \\
		\bottomrule
	\end{tabular}
\end{table}

Across Cases 1, 2, and 3, the observation time (per revisit) and number of revisits are changed while the total observation time is kept constant. Both the mean position and time error increase with fewer revisits, which is caused by a decrease in the light curve model fitting accuracy. Ideally, the light curve model should be fitted to a dense set of flux measurements evenly distributed in time. However, since the spacecraft must sequentially observe each star, the available flux measurements for any given star is highly segmented, such as those seen in Figure~\ref{fig:simulated-observations}. Longer observations of each star causes the time between revisits to increase, reducing the accuracy of the curve fit. Frequent revisits are more desirable, but this is limited by the capabilities of the spacecraft, such as whether there is enough propellant for many slew maneuvers.

In Cases 1, 4, and 5, the only variable changed is the exposure time, which affects the SNR of the flux measurements. In most cases, there is not much variation in the mean position and time errors, but the uncertainty increases due to uncertainty increases in the light curve model fits. The shortest exposure time case with MapCam is an exception and has much larger mean errors. 

In Cases 1, 6, and 7, the total observation time is varied by adjusting the number of revisits. If the total observation time is increased, more data is available to develop more accurate light curve fits, leading to decreased solution uncertainty. However, the spacecraft will have traveled a greater distance which is not accounted for by the navigation algorithm, so there can be a greater bias in the position and time solution. 

We emphasize that the present study only explores the impact of variations in broadly applicable observation parameters through the lens of OSIRIS-APEX. Other important factors, such as spacecraft trajectory and specific $\delta$ Scuti star selection, may also heavily affect navigation accuracy and must be assessed on a per-mission basis. Nonetheless, our findings suggest that autonomous recovery from being lost in space and time may be feasible using existing spacecraft instruments with our proposed method.

\section{Conclusions}\label{sec:conclusions}

A method for solving the lost in space and time problem using optical observations of $\delta$ Scuti stars is developed. The feasibility of the method is explored with consideration for detector performance constraints, $\delta$ Scuti star pulsation models, and spacecraft motion. Case studies indicate that it may be possible for OSIRIS-APEX to recover position and time information using onboard observations of $\delta$ Scuti stars alone to accuracies within \qty{0.03}{\au} (3$\sigma$) and \qty{3}{\second} (3$\sigma$), respectively.

\backmatter


%
%

\bmhead{Acknowledgements}

This paper includes data collected by the TESS mission. Funding for the TESS mission is provided by the NASA's Science Mission Directorate.

%
%

\noindent

%
%
%
%

\begin{appendices}

\section{Fixed Line-of-Sight Assumption}\label{appendix:los-assumption}

The constant LOS assumption impacts navigation accuracy by affecting Eqs.~\eqref{eqn:light-curve-observer} and \eqref{eqn:timing-eqn}. In the following derivation, suppose that the pointing error between the fixed LOS vector $\hat{\mb{u}}$ and true LOS vector $\hat{\mb{u}}^*$ is $\theta$. 

In Eq.~\eqref{eqn:light-curve-observer}, the timing of all flux measurements would be shifted by an amount bounded by $\epsilon_t$, which directly affects the resulting estimate of $T$ by the same amount:
\begin{equation}
	\epsilon_t = (\hat{\mb{u}}-\hat{\mb{u}}^*) \cdot (\mb{p}-\mb{b})/c
	\;\leq\; 2\sin(\theta/2) \norm{\mb{r}}/c
	\;<\; \theta \cdot{} \norm{\mb{r}}/c 
\end{equation}

Similarly in Eq.~\eqref{eqn:timing-eqn}, the error in the left-hand-side term $\hat{\mb{u}}\cdot{}\mb{p}$ can be equated to an error in $T$ on the right-hand-side:
\begin{equation}
	\epsilon_t = (\hat{\mb{u}}_i-\hat{\mb{u}}_i^*) \cdot \mb{p}/c
	\;\leq\; 2\sin(\theta/2) \norm{\mb{p}}/c
	\;<\; \theta \cdot \norm{\mb{p}}/c
\end{equation}

The $\delta$ Scuti stars considered in this study | shown in Table~\ref{tab:dsct-table} | have proper motion and annual parallaxes no greater than \qty{450}{\mas/yr} and \qty{130}{\mas} respectively, so the LOS pointing error $\theta$ is within $\qty{1}{\arcsecond} \approx \qty{5e-6}{\radian}$. Therefore, the error in $T$ introduced by the constant LOS assumption is on the order of \qty{2.5}{\ms} for $\norm{\mb{p}}$, $\norm{\mb{r}}$ on the order of \qty{1}{\au}, or \qty{0.25}{\second} at \qty{100}{\au}. For reference, errors in the estimate of $T$ are typically between 1 to 100 seconds in the cases studied in this work, which is much larger than the error introduced by the constant LOS assumption.

\section{Stability Analysis for \texorpdfstring{$\delta$}{Delta} Scuti Stars}\label{appendix:stability-analysis}

\begin{figure}[!ht]
	\centering
	\begin{subfigure}[t]{0.47\linewidth}
		\centering
		\includegraphics[width=\linewidth]{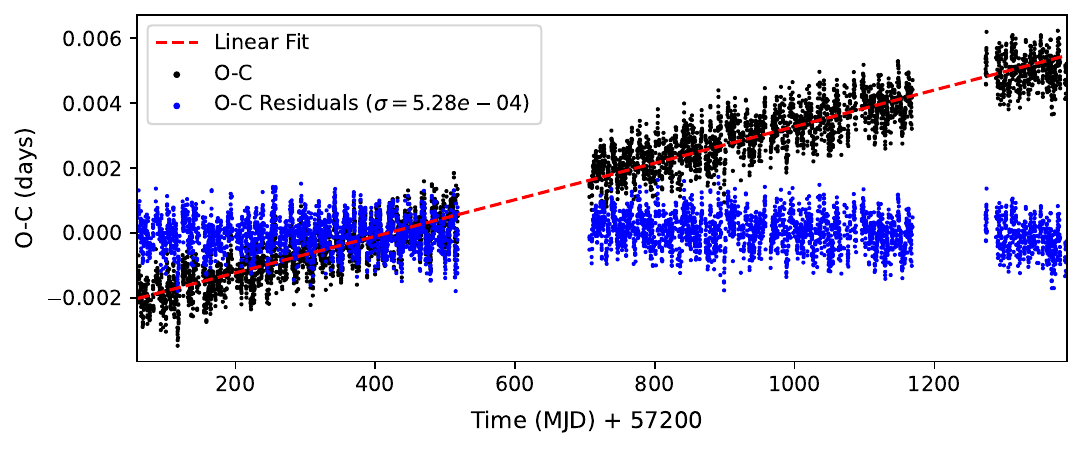}
		\caption{KIC 2297728}
	\end{subfigure}
	\hfill
	\begin{subfigure}[t]{0.47\linewidth}
		\centering
		\includegraphics[width=\linewidth]{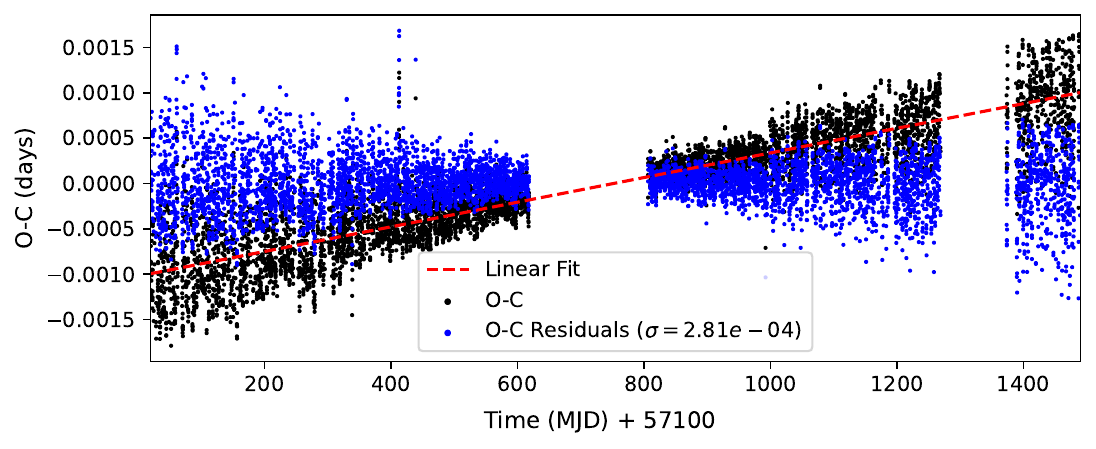}
		\caption{KIC 2304168}
	\end{subfigure}
	\begin{subfigure}[t]{0.47\linewidth}
		\centering
		\includegraphics[width=\linewidth]{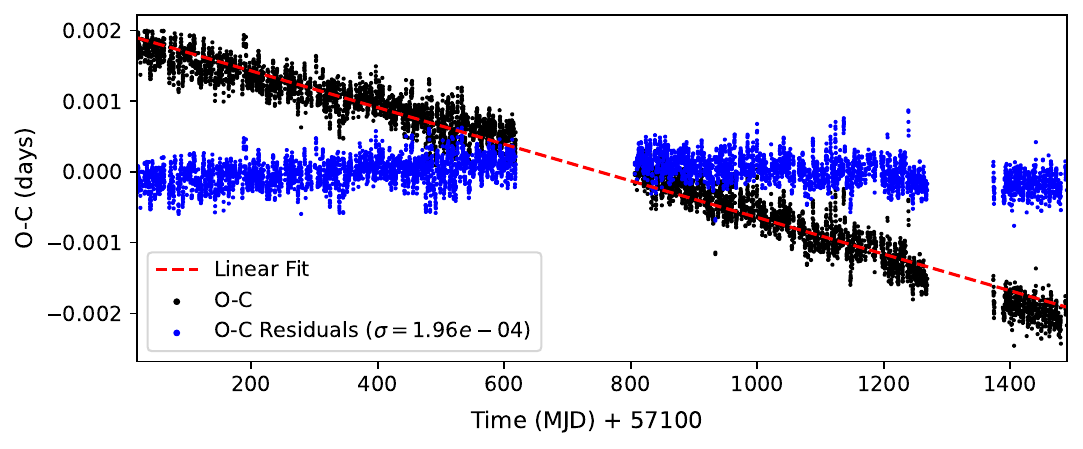}
		\caption{KIC 3123138}
	\end{subfigure}
	\hfill
	\begin{subfigure}[t]{0.47\linewidth}
		\centering
		\includegraphics[width=\linewidth]{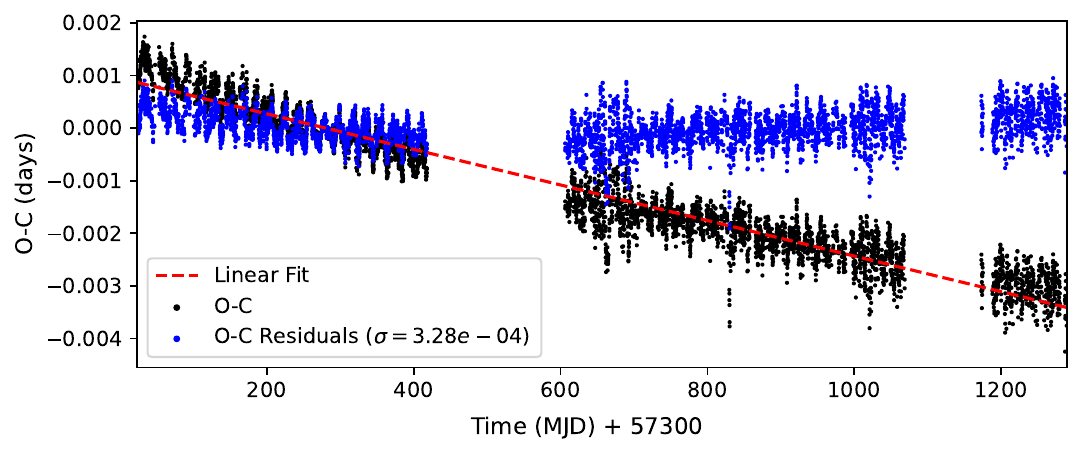}
		\caption{KIC 9353572}
	\end{subfigure}
	\caption{O-C diagrams for some $\delta$ Scuti stars with no amplitude modulation in the Kepler Input Catalog. }
	\label{fig:oc-diagrams}
\end{figure}

A previous study by Bowman et al. found 380 $\delta$ Scuti stars within the Kepler Input Catalog (38.7\%) with no significant amplitude modulation (meaning amplitude deviation less than $5\sigma$ from the mean over four years) \cite{bowman2016amplitude}. In this appendix, we plot the O-C diagrams for several stars listed in Ref.~\cite{bowman2016amplitude} in Figure~\ref{fig:oc-diagrams} to demonstrate their stability.

The simulated ground truth light curve model, which uses pulsation modes with SNR greater than 3 and amplitude greater than 1\% that of the highest amplitude mode, is fitted to Kepler observation data using Eq.~\eqref{eqn:objective-fcn} to determine the signal TOA. The calculated signal TOA is then compared to the observed TOA from Kepler using an O-C diagram. 

We note that the O-C residuals exhibit a linear trend with respect to time, which typically suggests frequency modulation in one or more of the stars' pulsation modes. Frequency modulation could be accounted for in a star's light curve model, though this is not considered in the present work. Detrending the data reveals O-C residuals (and therefore signal TOA errors) on the order of \qty{e-4}{\day} ($1\sigma$), which is between 10-100 seconds. Since signal TOA error is directly related to the position and time solution error by Eq.~\eqref{eqn:matrix-definitions}, this equates to $1\sigma$ position errors (along a single LOS) of 0.02-0.2 \qty{}{\au}.  

We did not perform an extensive analysis similar to that of Bowman et al. to verify the stability $\delta$ Scuti stars in the TESS catalog, but still used $\delta$ Scuti stars within TESS in our navigation case study. This is because stars from Kepler only span a small portion of the sky, which would lead to geometric dilution of precision in the navigation solution. Since 38.7\% of $\delta$ Scuti stars from Kepler were stable, we assume that is it equally feasible to identify stable $\delta$ Scuti stars over the entire celestial sphere.

\FloatBarrier
\section{\texorpdfstring{$\delta$}{Delta} Scuti Stars within Visibility Criteria}\label{secA1}

\renewcommand{\arraystretch}{0.8}
\begin{longtable}{ccccrr}
	\caption{List of $\delta$ Scuti stars with $m_V < 7.0$, $\Delta{V} > 0.04 \text{ mag}$, and $P < 0.20 \text{ d}$. $P$ is the period corresponding to the dominant pulsation mode frequency of the star. }\\
	\label{tab:dsct-table}\\
	\toprule
	Name & Max (Vmag) & Amplitude (Vmag) & Period (d) & \multicolumn{1}{c}{RA (deg)} & \multicolumn{1}{c}{DEC (deg)} \\
	\midrule
	\endfirsthead
	\caption*{\textbf{Table~\ref{tab:dsct-table} continued.}}\\
	\toprule
	Name & Max (Vmag) & Amplitude (Vmag) & Period (d) & \multicolumn{1}{c}{RA (deg)} & \multicolumn{1}{c}{DEC (deg)} \\
	\midrule
	\endhead
		V0474 Mon & 5.94 & 0.370 & 0.136 & 89.754 & -9.382 \\
		OX Aur & 5.94 & 0.200 & 0.154 & 103.256 & 38.869 \\
		del Sct & 4.60 & 0.190 & 0.194 & 280.568 & -9.053 \\
		rho Pup & 2.68 & 0.190 & 0.141 & 121.886 & -24.304 \\
		V0509 Per & 6.47 & 0.170 & 0.146 & 45.986 & 47.848 \\
		V0376 Per & 5.77 & 0.140 & 0.099 & 57.284 & 43.963 \\
		IM Tau & 5.33 & 0.130 & 0.145 & 62.708 & 26.481 \\
		ups UMa & 3.71 & 0.120 & 0.159 & 147.747 & 59.039 \\
		VW Ari & 6.64 & 0.120 & 0.161 & 36.690 & 10.565 \\
		VX Psc & 5.90 & 0.120 & 0.131 & 22.470 & 18.356 \\
		del Del & 4.38 & 0.110 & 0.157 & 310.865 & 15.075 \\
		bet Cep & 3.16 & 0.110 & 0.190 & 322.165 & 70.561 \\
		WZ Scl & 6.52 & 0.100 & 0.096 & 22.181 & -33.764 \\
		LT Vul & 6.52 & 0.100 & 0.109 & 285.927 & 21.268 \\
		VY Crt & 6.85 & 0.100 & 0.136 & 175.492 & -24.386 \\
		BV Cir & 6.80 & 0.100 & 0.158 & 225.259 & -64.576 \\
		X Cae & 6.28 & 0.100 & 0.135 & 76.109 & -35.705 \\
		tet Tuc & 6.06 & 0.090 & 0.049 & 8.347 & -71.266 \\
		CN Dra & 6.29 & 0.090 & 0.100 & 296.686 & 68.438 \\
		V0784 Cas & 6.61 & 0.090 & 0.109 & 32.606 & 59.980 \\
		AI CVn & 5.99 & 0.090 & 0.116 & 185.946 & 42.543 \\
		HT Peg & 5.30 & 0.090 & 0.060 & 358.155 & 10.947 \\
		FM Vir & 5.20 & 0.080 & 0.072 & 191.404 & 7.673 \\
		V0386 Per & 6.50 & 0.080 & 0.052 & 59.513 & 34.814 \\
		kap 2 Boo & 4.50 & 0.080 & 0.076 & 213.371 & 51.790 \\
		FM Com & 6.40 & 0.080 & 0.055 & 184.758 & 26.008 \\
		HH CMa & 6.59 & 0.070 & 0.190 & 104.312 & -22.203 \\
		WW LMi & 6.16 & 0.070 & 0.127 & 163.676 & 25.491 \\
		V0571 Mon & 5.43 & 0.070 & 0.089 & 107.848 & -0.302 \\
		AZ CMi & 6.44 & 0.070 & 0.095 & 116.032 & 2.405 \\
		eta Hya & 4.27 & 0.060 & 0.170 & 130.806 & 3.399 \\
		V0696 Tau & 5.22 & 0.060 & 0.036 & 65.151 & 15.095 \\
		gam CrB & 3.80 & 0.060 & 0.030 & 235.686 & 26.296 \\
		BT Cnc & 6.66 & 0.060 & 0.102 & 129.928 & 19.778 \\
		CC Gru & 6.62 & 0.060 & 0.126 & 339.785 & -52.692 \\
		rho Phe & 5.20 & 0.060 & 0.120 & 12.672 & -50.987 \\
		eps Cep & 4.15 & 0.060 & 0.041 & 333.759 & 57.044 \\
		rho Pav & 4.85 & 0.055 & 0.114 & 309.397 & -61.530 \\
		UU Ari & 6.10 & 0.050 & 0.068 & 37.660 & 19.855 \\
		GN And & 5.23 & 0.050 & 0.069 & 7.531 & 29.752 \\
		gam UMi & 3.04 & 0.050 & 0.143 & 230.182 & 71.834 \\
		sig Oct & 5.45 & 0.050 & 0.097 & 317.195 & -88.957 \\
		V1644 Cyg & 4.92 & 0.050 & 0.031 & 303.633 & 36.806 \\
		FG Vir & 6.53 & 0.050 & 0.079 & 183.564 & -5.717 \\
		DK Vir & 6.67 & 0.050 & 0.119 & 199.106 & -1.391 \\
		V0637 Mon & 4.96 & 0.050 & 0.191 & 105.728 & -4.239 \\
		del Cet & 4.05 & 0.050 & 0.161 & 39.871 & 0.329 \\
		EN UMa & 5.83 & 0.050 & 0.155 & 155.264 & 68.748 \\
		V1208 Aql & 5.51 & 0.050 & 0.150 & 289.914 & 12.375 \\
		iot Boo & 4.73 & 0.050 & 0.027 & 214.041 & 51.367 \\
		AO CVn & 4.70 & 0.050 & 0.122 & 199.386 & 40.573 \\
		omi 1 Eri & 4.00 & 0.050 & 0.082 & 62.966 & -6.838 \\
		alf Mus & 2.68 & 0.050 & 0.090 & 189.296 & -69.136 \\
		bet Cen & 0.61 & 0.045 & 0.157 & 210.956 & -60.373 \\
		del Lup & 3.20 & 0.040 & 0.165 & 230.343 & -40.648 \\
		V0620 Her & 6.19 & 0.040 & 0.080 & 257.763 & 24.238 \\
		HD 46089 & 5.21 & 0.040 & 0.050 & 97.951 & 11.544 \\
		tau 1 Lup & 4.54 & 0.040 & 0.177 & 216.534 & -45.221 \\
		UV Ari & 5.18 & 0.040 & 0.035 & 41.240 & 12.446 \\
		FT Vir & 6.20 & 0.040 & 0.050 & 186.965 & -4.615 \\
		V0644 Her & 6.32 & 0.040 & 0.115 & 253.817 & 13.620 \\
		bet Cas & 2.25 & 0.040 & 0.101 & 2.295 & 59.150 \\
		V0647 Pup & 6.74 & 0.040 & 0.117 & 91.946 & -44.729 \\
		HQ Hya & 6.29 & 0.040 & 0.075 & 124.813 & -10.166 \\
		V0480 Tau & 5.09 & 0.040 & 0.042 & 72.844 & 18.840 \\
		VV Ari & 6.69 & 0.040 & 0.076 & 27.789 & 20.514 \\		
	\bottomrule
\end{longtable}





\end{appendices}


\bibliography{references}


\begin{thebibliography}{31}
\ifx \bisbn   \undefined \def \bisbn  #1{ISBN #1}\fi
\ifx \binits  \undefined \def \binits#1{#1}\fi
\ifx \bauthor  \undefined \def \bauthor#1{#1}\fi
\ifx \batitle  \undefined \def \batitle#1{#1}\fi
\ifx \bjtitle  \undefined \def \bjtitle#1{#1}\fi
\ifx \bvolume  \undefined \def \bvolume#1{\textbf{#1}}\fi
\ifx \byear  \undefined \def \byear#1{#1}\fi
\ifx \bissue  \undefined \def \bissue#1{#1}\fi
\ifx \bfpage  \undefined \def \bfpage#1{#1}\fi
\ifx \blpage  \undefined \def \blpage #1{#1}\fi
\ifx \burl  \undefined \def \burl#1{\textsf{#1}}\fi
\ifx \doiurl  \undefined \def \doiurl#1{\url{https://doi.org/#1}}\fi
\ifx \betal  \undefined \def \betal{\textit{et al.}}\fi
\ifx \binstitute  \undefined \def \binstitute#1{#1}\fi
\ifx \binstitutionaled  \undefined \def \binstitutionaled#1{#1}\fi
\ifx \bctitle  \undefined \def \bctitle#1{#1}\fi
\ifx \beditor  \undefined \def \beditor#1{#1}\fi
\ifx \bpublisher  \undefined \def \bpublisher#1{#1}\fi
\ifx \bbtitle  \undefined \def \bbtitle#1{#1}\fi
\ifx \bedition  \undefined \def \bedition#1{#1}\fi
\ifx \bseriesno  \undefined \def \bseriesno#1{#1}\fi
\ifx \blocation  \undefined \def \blocation#1{#1}\fi
\ifx \bsertitle  \undefined \def \bsertitle#1{#1}\fi
\ifx \bsnm \undefined \def \bsnm#1{#1}\fi
\ifx \bsuffix \undefined \def \bsuffix#1{#1}\fi
\ifx \bparticle \undefined \def \bparticle#1{#1}\fi
\ifx \barticle \undefined \def \barticle#1{#1}\fi
\bibcommenthead
\ifx \bconfdate \undefined \def \bconfdate #1{#1}\fi
\ifx \botherref \undefined \def \botherref #1{#1}\fi
\ifx \url \undefined \def \url#1{\textsf{#1}}\fi
\ifx \bchapter \undefined \def \bchapter#1{#1}\fi
\ifx \bbook \undefined \def \bbook#1{#1}\fi
\ifx \bcomment \undefined \def \bcomment#1{#1}\fi
\ifx \oauthor \undefined \def \oauthor#1{#1}\fi
\ifx \citeauthoryear \undefined \def \citeauthoryear#1{#1}\fi
\ifx \endbibitem  \undefined \def \endbibitem {}\fi
\ifx \bconflocation  \undefined \def \bconflocation#1{#1}\fi
\ifx \arxivurl  \undefined \def \arxivurl#1{\textsf{#1}}\fi
\csname PreBibitemsHook\endcsname

\bibitem[\protect\citeauthoryear{Fountain et~al.}{2008}]{fountain2008new}
\begin{barticle}
\bauthor{\bsnm{Fountain}, \binits{G.H.}},
\bauthor{\bsnm{Kusnierkiewicz}, \binits{D.Y.}},
\bauthor{\bsnm{Hersman}, \binits{C.B.}},
\bauthor{\bsnm{Herder}, \binits{T.S.}},
\bauthor{\bsnm{Coughlin}, \binits{T.B.}},
\bauthor{\bsnm{Gibson}, \binits{W.C.}},
\bauthor{\bsnm{Clancy}, \binits{D.A.}},
\bauthor{\bsnm{DeBoy}, \binits{C.C.}},
\bauthor{\bsnm{Hill}, \binits{T.A.}},
\bauthor{\bsnm{Kinnison}, \binits{J.D.}}, \betal:
\batitle{{The New Horizons Spacecraft}}.
\bjtitle{Space science reviews}
\bvolume{140},
\bfpage{23}--\blpage{47}
(\byear{2008})
\doiurl{10.1007/s11214-008-9374-8}
\end{barticle}
\endbibitem

\bibitem[\protect\citeauthoryear{Ecale et~al.}{2018}]{ecale2018juice}
\begin{bchapter}
\bauthor{\bsnm{Ecale}, \binits{E.}},
\bauthor{\bsnm{Torelli}, \binits{F.}},
\bauthor{\bsnm{Tanco}, \binits{I.}}:
\bctitle{{JUICE interplanetary operations design: drivers and challenges}}.
In: \bbtitle{2018 SpaceOps Conference},
p. \bfpage{2493}
(\byear{2018}).
\doiurl{10.2514/6.2018-2493}
\end{bchapter}
\endbibitem

\bibitem[\protect\citeauthoryear{Keil and Herfort}{2007}]{keil2007contingency}
\begin{bchapter}
\bauthor{\bsnm{Keil}, \binits{J.}},
\bauthor{\bsnm{Herfort}, \binits{U.}}:
\bctitle{{Contingency Operations during Failure of Inertial Attitude Acquisition Due to Star Tracker Blinding for Three-Axes-Stabilized Interplanetary Spacecraft}}.
In: \bbtitle{Proceedings of the 20th International Symposium on Space Flight Dynamics}
(\byear{2007}).
\bcomment{\url{https://ntrs.nasa.gov/citations/20080012644}}
\end{bchapter}
\endbibitem

\bibitem[\protect\citeauthoryear{Cox and Ossing}{2018}]{cox2018fall}
\begin{bchapter}
\bauthor{\bsnm{Cox}, \binits{M.W.}},
\bauthor{\bsnm{Ossing}, \binits{D.A.}}:
\bctitle{The fall and rise of stereo behind}.
In: \bbtitle{2018 SpaceOps Conference},
p. \bfpage{2565}
(\byear{2018}).
\doiurl{10.2514/6.2018-2565}
\end{bchapter}
\endbibitem

\bibitem[\protect\citeauthoryear{{Jet Propulsion Laboratory}}{2023}]{voyager2news}
\begin{botherref}
\oauthor{\bsnm{{Jet Propulsion Laboratory}}}:
{NASA} Mission Update: Voyager 2 Communications Pause.
\url{https://www.jpl.nasa.gov/news/nasa-mission-update-voyager-2-communications-pause} [Accessed Jan. 10, 2024]
(2023)
\end{botherref}
\endbibitem

\bibitem[\protect\citeauthoryear{Johnston}{2020}]{johnston2020scheduling}
\begin{bchapter}
\bauthor{\bsnm{Johnston}, \binits{M.D.}}:
\bctitle{{Scheduling NASA's Deep Space Network: Priorities, Preferences, and Optimization}}.
In: \bbtitle{ICAPS - SPARK Workshop}
(\byear{2020}).
\bcomment{JPL Open Repository, \url{https://hdl.handle.net/2014/53269}}
\end{bchapter}
\endbibitem

\bibitem[\protect\citeauthoryear{Tanygin}{2014}]{tanygin2014closed}
\begin{barticle}
\bauthor{\bsnm{Tanygin}, \binits{S.}}:
\batitle{Closed-form solution for lost-in-space visual navigation problem}.
\bjtitle{Journal of Guidance, Control, and Dynamics}
\bvolume{37}(\bissue{6}),
\bfpage{1754}--\blpage{1766}
(\byear{2014})
\doiurl{10.2514/1.G000529}
\end{barticle}
\endbibitem

\bibitem[\protect\citeauthoryear{Hollenberg and Christian}{2020}]{hollenberg2020geometric}
\begin{barticle}
\bauthor{\bsnm{Hollenberg}, \binits{C.L.}},
\bauthor{\bsnm{Christian}, \binits{J.A.}}:
\batitle{Geometric solutions for problems in velocity-based orbit determination}.
\bjtitle{The Journal of the Astronautical Sciences}
\bvolume{67}(\bissue{1}),
\bfpage{188}--\blpage{224}
(\byear{2020})
\doiurl{10.1007/s40295-019-00170-7}
\end{barticle}
\endbibitem

\bibitem[\protect\citeauthoryear{Hou and Putnam}{2024}]{hou2022xnav}
\begin{barticle}
\bauthor{\bsnm{Hou}, \binits{L.}},
\bauthor{\bsnm{Putnam}, \binits{Z.R.}}:
\batitle{A norm-minimization algorithm for solving the lost-in-space problem with xnav}.
\bjtitle{The Journal of the Astronautical Sciences}
\bvolume{71}(\bissue{1}),
\bfpage{6}
(\byear{2024})
\doiurl{10.1007/s40295-023-00425-4}
\end{barticle}
\endbibitem

\bibitem[\protect\citeauthoryear{Adams and Peck}{2017}]{adams2017lost}
\begin{bchapter}
\bauthor{\bsnm{Adams}, \binits{V.H.}},
\bauthor{\bsnm{Peck}, \binits{M.A.}}:
\bctitle{Lost in space and time}.
In: \bbtitle{AIAA Guidance, Navigation, and Control Conference},
p. \bfpage{1030}
(\byear{2017}).
\doiurl{10.2514/6.2017-1030}
\end{bchapter}
\endbibitem

\bibitem[\protect\citeauthoryear{Dahir}{2020}]{dahir2020lost}
\begin{botherref}
\oauthor{\bsnm{Dahir}, \binits{A.R.}}:
Lost in space: Autonomous deep space navigation.
PhD thesis,
University of Colorado at Boulder
(2020).
\url{https://www.proquest.com/docview/2408808776}
\end{botherref}
\endbibitem

\bibitem[\protect\citeauthoryear{Sala~{\'A}lvarez et~al.}{2004}]{sala2004feasibility}
\begin{botherref}
\oauthor{\bsnm{Sala~{\'A}lvarez}, \binits{J.}},
\oauthor{\bsnm{Urruela~Planas}, \binits{A.}},
\oauthor{\bsnm{Villares~Piera}, \binits{N.J.}}:
Feasibility study for a spacecraft navigation system relying on pulsar timing information
(2004).
\url{https://upcommons.upc.edu/bitstream/handle/2117/11514/final_report_23062004.pdf}
\end{botherref}
\endbibitem

\bibitem[\protect\citeauthoryear{Broschart et~al.}{2019}]{broschart2019kinematic}
\begin{barticle}
\bauthor{\bsnm{Broschart}, \binits{S.B.}},
\bauthor{\bsnm{Bradley}, \binits{N.}},
\bauthor{\bsnm{Bhaskaran}, \binits{S.}}:
\batitle{Kinematic approximation of position accuracy achieved using optical observations of distant asteroids}.
\bjtitle{Journal of Spacecraft and Rockets}
\bvolume{56}(\bissue{5}),
\bfpage{1383}--\blpage{1392}
(\byear{2019})
\doiurl{10.2514/1.A34354}
\end{barticle}
\endbibitem

\bibitem[\protect\citeauthoryear{Eyer and Mowlavi}{2008}]{eyer2008variable}
\begin{bchapter}
\bauthor{\bsnm{Eyer}, \binits{L.}},
\bauthor{\bsnm{Mowlavi}, \binits{N.}}:
\bctitle{Variable stars across the observational hr diagram}.
In: \bbtitle{Journal of Physics: Conference Series},
vol. \bseriesno{118},
p. \bfpage{012010}
(\byear{2008}).
\doiurl{10.1088/1742-6596/118/1/012010} .
\bcomment{IOP Publishing}
\end{bchapter}
\endbibitem

\bibitem[\protect\citeauthoryear{Aerts et~al.}{2010}]{aerts2010asteroseismology}
\begin{bchapter}
\bauthor{\bsnm{Aerts}, \binits{C.}},
\bauthor{\bsnm{Christensen-Dalsgaard}, \binits{J.}},
\bauthor{\bsnm{Kurtz}, \binits{D.W.}}:
\bctitle{3}.
\bbtitle{Theory of Stellar Oscillations}.
\bpublisher{{Springer Dordrecht}},
\blocation{Netherlands}
(\byear{2010}).
\doiurl{10.1007/978-1-4020-5803-5}
\end{bchapter}
\endbibitem

\bibitem[\protect\citeauthoryear{Handler}{2009}]{handler2009delta}
\begin{bchapter}
\bauthor{\bsnm{Handler}, \binits{G.}}:
\bctitle{Delta scuti variables}.
In: \bbtitle{AIP Conference Proceedings},
vol. \bseriesno{1170},
pp. \bfpage{403}--\blpage{409}
(\byear{2009}).
\doiurl{10.1063/1.3246528} .
\bcomment{American Institute of Physics}
\end{bchapter}
\endbibitem

\bibitem[\protect\citeauthoryear{Bowman}{2017}]{bowman2017amplitude}
\begin{bbook}
\bauthor{\bsnm{Bowman}, \binits{D.M.}}:
\bbtitle{Amplitude Modulation of Pulsation Modes in Delta Scuti Stars}.
\bpublisher{{Springer Cham}},
\blocation{Switzerland}
(\byear{2017}).
\doiurl{10.1007/978-3-319-66649-5}
\end{bbook}
\endbibitem

\bibitem[\protect\citeauthoryear{Breger}{2009}]{breger2009period}
\begin{bchapter}
\bauthor{\bsnm{Breger}, \binits{M.}}:
\bctitle{Period variations of delta scuti stars}.
In: \bbtitle{AIP Conference Proceedings},
vol. \bseriesno{1170},
pp. \bfpage{410}--\blpage{414}
(\byear{2009}).
\doiurl{10.1063/1.3246530} .
\bcomment{American Institute of Physics}
\end{bchapter}
\endbibitem

\bibitem[\protect\citeauthoryear{Mantegazza et~al.}{1999}]{mantegazza1999simultaneous}
\begin{barticle}
\bauthor{\bsnm{Mantegazza}, \binits{L.}},
\bauthor{\bsnm{Zerbi}, \binits{F.M.}},
\bauthor{\bsnm{Sacchi}, \binits{A.}}:
\batitle{Simultaneous intensive photometry and high resolution spectroscopy of delta scuti stars iv. an improved picture of the pulsational behaviour of x caeli}.
\bjtitle{arXiv preprint astro-ph/9911337}
(\byear{1999})
\doiurl{10.48550/arXiv.astro-ph/9911337}
\end{barticle}
\endbibitem

\bibitem[\protect\citeauthoryear{Kovacs and Paparo}{1989}]{kovacs1989photoelectric}
\begin{barticle}
\bauthor{\bsnm{Kovacs}, \binits{G.}},
\bauthor{\bsnm{Paparo}, \binits{M.}}:
\batitle{Photoelectric observations and analysis of the low-amplitude delta scuti star 78 tau}.
\bjtitle{Monthly Notices of the Royal Astronomical Society}
\bvolume{237}(\bissue{1}),
\bfpage{201}--\blpage{217}
(\byear{1989})
\doiurl{10.1093/mnras/237.1.201}
\end{barticle}
\endbibitem

\bibitem[\protect\citeauthoryear{Poretti}{2001}]{poretti2001fourier}
\begin{barticle}
\bauthor{\bsnm{Poretti}, \binits{E.}}:
\batitle{Fourier decomposition and frequency analysis of the pulsating stars with 1 d in the ogle database-i. monoperiodic delta scuti, rrc and rrab variables. separation criteria and particularities}.
\bjtitle{Astronomy \& Astrophysics}
\bvolume{371}(\bissue{3}),
\bfpage{986}--\blpage{996}
(\byear{2001})
\doiurl{10.1051/0004-6361:20010462}
\end{barticle}
\endbibitem

\bibitem[\protect\citeauthoryear{Zima}{2006}]{zima2006new}
\begin{barticle}
\bauthor{\bsnm{Zima}, \binits{W.}}:
\batitle{A new method for the spectroscopic identification of stellar non-radial pulsation modes-i. the method and numerical tests}.
\bjtitle{Astronomy \& Astrophysics}
\bvolume{455}(\bissue{1}),
\bfpage{227}--\blpage{234}
(\byear{2006})
\doiurl{10.1051/0004-6361:20064876}
\end{barticle}
\endbibitem

\bibitem[\protect\citeauthoryear{Bowman et~al.}{2016}]{bowman2016amplitude}
\begin{barticle}
\bauthor{\bsnm{Bowman}, \binits{D.M.}},
\bauthor{\bsnm{Kurtz}, \binits{D.W.}},
\bauthor{\bsnm{Breger}, \binits{M.}},
\bauthor{\bsnm{Murphy}, \binits{S.J.}},
\bauthor{\bsnm{Holdsworth}, \binits{D.L.}}:
\batitle{Amplitude modulation in $\delta$ sct stars: statistics from an ensemble study of kepler targets}.
\bjtitle{Monthly Notices of the Royal Astronomical Society}
\bvolume{460}(\bissue{2}),
\bfpage{1970}--\blpage{1989}
(\byear{2016})
\doiurl{10.1093/mnras/stw1153}
\end{barticle}
\endbibitem

\bibitem[\protect\citeauthoryear{Zhou and Jiang}{2011}]{zhou2011period}
\begin{barticle}
\bauthor{\bsnm{Zhou}, \binits{A.-Y.}},
\bauthor{\bsnm{Jiang}, \binits{S.-Y.}}:
\batitle{Period and amplitude variability of the high-amplitude $\delta$ scuti star gp andromedae}.
\bjtitle{The Astronomical Journal}
\bvolume{142}(\bissue{4}),
\bfpage{100}
(\byear{2011})
\doiurl{10.1088/0004-6256/142/4/100}
\end{barticle}
\endbibitem

\bibitem[\protect\citeauthoryear{Boonyarak et~al.}{2011}]{boonyarak2011period}
\begin{barticle}
\bauthor{\bsnm{Boonyarak}, \binits{C.}},
\bauthor{\bsnm{Fu}, \binits{J.-N.}},
\bauthor{\bsnm{Khokhuntod}, \binits{P.}},
\bauthor{\bsnm{Jiang}, \binits{S.-Y.}}:
\batitle{On the period variations of several low declination high amplitude delta scuti variables}.
\bjtitle{Astrophysics and Space Science}
\bvolume{333},
\bfpage{125}--\blpage{131}
(\byear{2011})
\doiurl{10.1007/s10509-010-0574-9}
\end{barticle}
\endbibitem

\bibitem[\protect\citeauthoryear{Sheikh et~al.}{2011}]{sheikh2011spacecraft}
\begin{barticle}
\bauthor{\bsnm{Sheikh}, \binits{S.I.}},
\bauthor{\bsnm{Hanson}, \binits{J.E.}},
\bauthor{\bsnm{Graven}, \binits{P.H.}},
\bauthor{\bsnm{Pines}, \binits{D.J.}}:
\batitle{Spacecraft navigation and timing using x-ray pulsars}.
\bjtitle{Navigation}
\bvolume{58}(\bissue{2}),
\bfpage{165}--\blpage{186}
(\byear{2011})
\doiurl{10.1002/j.2161-4296.2011.tb01799.x}
\end{barticle}
\endbibitem

\bibitem[\protect\citeauthoryear{Lohan}{2021}]{lohan2021methodology}
\begin{botherref}
\oauthor{\bsnm{Lohan}, \binits{K.}}:
Methodology for state determination without prior information using x-ray pulsar navigation systems.
PhD thesis,
University of Illinois at Urbana-Champaign
(2021).
\url{https://hdl.handle.net/2142/113068}
\end{botherref}
\endbibitem

\bibitem[\protect\citeauthoryear{DellaGiustina et~al.}{2023}]{dellagiustina2023osiris}
\begin{barticle}
\bauthor{\bsnm{DellaGiustina}, \binits{D.N.}},
\bauthor{\bsnm{Nolan}, \binits{M.C.}},
\bauthor{\bsnm{Polit}, \binits{A.T.}},
\bauthor{\bsnm{Moreau}, \binits{M.C.}},
\bauthor{\bsnm{Golish}, \binits{D.R.}},
\bauthor{\bsnm{Simon}, \binits{A.A.}},
\bauthor{\bsnm{Adam}, \binits{C.D.}},
\bauthor{\bsnm{Antreasian}, \binits{P.G.}},
\bauthor{\bsnm{Ballouz}, \binits{R.-L.}},
\bauthor{\bsnm{Barnouin}, \binits{O.S.}}, \betal:
\batitle{Osiris-apex: an osiris-rex extended mission to asteroid apophis}.
\bjtitle{The Planetary Science Journal}
\bvolume{4}(\bissue{10}),
\bfpage{198}
(\byear{2023})
\doiurl{10.3847/PSJ/acf75e}
\end{barticle}
\endbibitem

\bibitem[\protect\citeauthoryear{Rizk et~al.}{2018}]{rizk2018ocams}
\begin{barticle}
\bauthor{\bsnm{Rizk}, \binits{B.}},
\bauthor{\bsnm{Drouet~d’Aubigny}, \binits{C.}},
\bauthor{\bsnm{Golish}, \binits{D.}},
\bauthor{\bsnm{Fellows}, \binits{C.}},
\bauthor{\bsnm{Merrill}, \binits{C.}},
\bauthor{\bsnm{Smith}, \binits{P.}},
\bauthor{\bsnm{Walker}, \binits{M.}},
\bauthor{\bsnm{Hendershot}, \binits{J.}},
\bauthor{\bsnm{Hancock}, \binits{J.}},
\bauthor{\bsnm{Bailey}, \binits{S.}}, \betal:
\batitle{Ocams: the osiris-rex camera suite}.
\bjtitle{Space Science Reviews}
\bvolume{214},
\bfpage{1}--\blpage{55}
(\byear{2018})
\doiurl{10.1007/s11214-017-0460-7}
\end{barticle}
\endbibitem

\bibitem[\protect\citeauthoryear{Golish et~al.}{2020}]{golish2020ground}
\begin{barticle}
\bauthor{\bsnm{Golish}, \binits{D.}},
\bauthor{\bsnm{Drouet~d’Aubigny}, \binits{C.}},
\bauthor{\bsnm{Rizk}, \binits{B.}},
\bauthor{\bsnm{DellaGiustina}, \binits{D.}},
\bauthor{\bsnm{Smith}, \binits{P.}},
\bauthor{\bsnm{Becker}, \binits{K.}},
\bauthor{\bsnm{Shultz}, \binits{N.}},
\bauthor{\bsnm{Stone}, \binits{T.}},
\bauthor{\bsnm{Barker}, \binits{M.}},
\bauthor{\bsnm{Mazarico}, \binits{E.}}, \betal:
\batitle{Ground and in-flight calibration of the osiris-rex camera suite}.
\bjtitle{Space science reviews}
\bvolume{216},
\bfpage{1}--\blpage{31}
(\byear{2020})
\doiurl{10.1007/s11214-019-0626-6}
\end{barticle}
\endbibitem

\bibitem[\protect\citeauthoryear{Watson et~al.}{2006}]{watson2006international}
\begin{bchapter}
\bauthor{\bsnm{Watson}, \binits{C.L.}},
\bauthor{\bsnm{Henden}, \binits{A.A.}},
\bauthor{\bsnm{Price}, \binits{A.}}:
\bctitle{The international variable star index (vsx)}.
In: \bbtitle{The Society for Astronomical Sciences 25th Annual Symposium on Telescope Science. Held May 23-25, 2006, at Big Bear, CA. Published by the Society for Astronomical Sciences., P. 47},
vol. \bseriesno{25},
p. \bfpage{47}
(\byear{2006})
\end{bchapter}
\endbibitem

\end{thebibliography}

\end{document}